\documentclass[nofootinbib,floats,aps,superscriptaddress,preprint]{revtex4} 

\newcommand{\fnref}[1]{~\ref{#1}}
\def\gev{~{\rm GeV}}
\def\lthdm{\Lambda_{\rm 2HDM}}
\def\lam{\lambda}
\def\lamT{\lambda_T}
\def\lamU{\lambda_U}
\def\lamV{\lambda_V}
\def\lamA{\lambda_A}
\def\lamF{\lambda_F}
\def\wtil{\widetilde}
\def\ben{\begin{enumerate}}
\def\een{\end{enumerate}}
\def\beq{\begin{equation}}
\def\eeq{\end{equation}}
\def\beqa{\begin{eqnarray}}
\def\eeqa{\end{eqnarray}}
\def\crr{\crcr\noalign{\vskip .1in}}
\def\ifmath#1{\relax\ifmmode #1\else $#1$\fi}
\def\lsim{\mathrel{\raise.3ex\hbox{$<$\kern-.75em\lower1ex\hbox{$\sim$}}}}
\def\gsim{\mathrel{\raise.3ex\hbox{$>$\kern-.75em\lower1ex\hbox{$\sim$}}}}
\def\eq#1{eq.~(\ref{#1})}
\def\Ref#1{ref.~\cite{#1}}
\def\Refs#1#2{refs.~\cite{#1} and \cite{#2}}

\def\eqs#1#2{eqs.~(\ref{#1})--(\ref{#2})}
\def\Eq#1{Eq.~(\ref{#1})}
\def\Eqs#1#2{Eqs.~(\ref{#1})--(\ref{#2})}
\def\eqns#1#2{eqs.~(\ref{#1}) and (\ref{#2})}
\def\Eqns#1#2{Eqs.~(\ref{#1}) and (\ref{#2})}

\def\anti{\overline}
\def\rtwo{\sqrt 2}
\def\mud{M_U}
\def\mdd{M_D}
\def\vev#1{\langle #1 \rangle}
\def\qlo{Q^0_L}
\def\uro{U^0_R}
\def\dro{D^0_R}

\def\eiuo{\eta_1^{U,0}}
\def\eiiuo{\eta_2^{U,0}}
\def\eido{\eta_1^{D,0}}
\def\eiido{\eta_2^{D,0}}
\def\eiu{\eta_1^U}
\def\eiiu{\eta_2^U}
\def\eid{\eta_1^D}
\def\eiid{\eta_2^D}
\def\eiuoi{\eta_i^{U,0}}
\def\eidoi{\eta_i^{D,0}}
\def\eiui{\eta_i^U}
\def\eidi{\eta_i^D}
\def\Lam{\Lambda}
\def\lambar{\lam}
\def\calb{\mathcal{B}}
\def\calo{\mathcal{O}}
\def\cbpa{c_{\beta+\alpha}}
\def\cbma{c_{\beta-\alpha}}
\def\sbpa{s_{\beta+\alpha}}
\def\sbma{s_{\beta-\alpha}}
\def\ctwob{c_{2\beta}}
\def\stwob{s_{2\beta}}
\def\sthreeb{s_{3\beta}}
\def\sthreea{s_{3\alpha}}
\def\cthreeb{c_{3\beta}}
\def\ctwoa{c_{2\alpha}}
\def\cthreea{c_{3\alpha}}
\def\stwoa{s_{2\alpha}}
\def\lamtil{\lam\ls{345}}
\def\lamhat{\widehat\lam}
\def\phm{\phantom{-}}

\def\beq{\begin{equation}}
\def\eeq{\end{equation}}
\def\ifmath#1{\relax\ifmmode #1\else $#1$\fi}
\def\calm{\mathcal{M}}
\def\calv{\mathcal{V}}
\def\call{\mathcal{L}}
\def\tb{t_{\beta}}
\def\tbi{t_\beta^{-1}}

\def\sb  {s_{\beta}}
\def\cb  {c_{\beta}}
\def\stwob  {s_{2\beta}}
\def\ctwob  {c_{2\beta}}
\def\sa  {s_{\alpha}}
\def\ca  {c_{\alpha}}

\def\cab  {c_{\alpha+\beta}}
\def\sba  {s_{\beta-\alpha}}
\def\cba  {c_{\beta-\alpha}}
\def\tanb{\tan\beta}
\def\cotb{\cot\beta}
\def\sinb{\sin\beta}
\def\cosb{\cos\beta}

\def\sinbma{\sin(\beta-\alpha)}
\def\cosbma{\cos(\beta-\alpha)}
\def\sbmaii{s^2_{\beta-\alpha}}
\def\cbmaii{c^2_{\beta-\alpha}}

\def\hsm{h_{\rm SM}}

\def\hl{h}
\def\ha{A}
\def\hh{H}
\def\hpm{{H^\pm}}
\def\gpm{{G^\pm}}
\def\gmp{{G^\mp}}

\def\wpm{{W^\pm}}
\def\hp{{H^+}}
\def\hm{{H^-}}
\def\go{G}
\def\g{G}
\def\gp{{G^+}}
\def\gm{{G^-}}
\def\lamtil{\lam_{345}}
\def\nncr{\nonumber\\[5pt]}
\def\nncrsmall{\nonumber\cr}
\def\mha{m_{\ha}}
\def\mhl{m_{\hl}}
\def\mhh{m_{\hh}}
\def\mhpm{m_{\hpm}}
\def\mz{m_Z}
\def\mw{m_W}

\def\ls#1{\ifmath{_{\lower1.5pt\hbox{$\scriptstyle #1$}}}}
\def\lss#1{\ifmath{^{\,\lower2.5pt\hbox{$\scriptstyle #1$}}}}
\def\nicefrac#1#2{\hbox{${#1\over #2}$}}
\def\ihalf{\nicefrac{i}{2}}
\def\half{\ifmath{{\textstyle{1 \over 2}}}}

\def\thalf{\ifmath{{\textstyle{3 \over 2}}}}
\def\quarter{\ifmath{{\textstyle{1 \over 4}}}}
\def\eighth{\ifmath{{\textstyle{1 \over 8}}}}
\def\mlsq{m\ls{L}^2}
\def\mtsq{m\ls{T}^2}

\def\mssq{m\ls{S}^2}

\def\ie{{\it i.e.}}
\def\eg{{\it e.g.}}

\def\vs{{\it vs.}}

\renewcommand{\Re}{\rm Re}
\renewcommand{\Im}{\rm Im}

\begin{document}
%
\preprint{
\vbox{
      \hbox{UCD-2002-10}
      \hbox{SCIPP-02/10}
      \hbox{hep-ph/0207010}
      \hbox{July 2002}
    }}
\vspace*{1cm}  

\title{The CP-conserving two-Higgs-doublet model: 
 the approach to the decoupling limit}
\author{John F. Gunion}
\affiliation{ Davis Institute for High Energy Physics \\
University of California, Davis, CA 95616, U.S.A.}

\author{Howard E. Haber}
\affiliation{Santa Cruz Institute for Particle Physics  \\
   University of California, Santa Cruz, CA 95064, U.S.A.
\\[20pt] $\phantom{}$ }

\begin{abstract}
A CP-even neutral Higgs boson with Standard-Model-like couplings may
be the lightest scalar of a two-Higgs-doublet model.
We study the decoupling limit of the most general 
CP-conserving two-Higgs-doublet
model, where the mass of the lightest Higgs scalar is significantly
smaller than the masses of the other Higgs bosons of the model.
In this case, the properties of the lightest Higgs boson are nearly
indistinguishable from those of the Standard Model Higgs boson.
The first non-trivial corrections to Higgs couplings in the 
approach to the decoupling limit are also evaluated.  The importance
of detecting such deviations in precision Higgs measurements at future
colliders is emphasized.
We also clarify the case in which a neutral Higgs boson can
possess Standard-Model-like couplings in a regime where the decoupling
limit does not apply.   The two-Higgs-doublet sector of the minimal
supersymmetric model illustrates many of the above features. 

\end{abstract}

\maketitle

\section{Introduction}  \label{sec:intro}

The minimal version of the Standard Model (SM) contains one
complex Higgs doublet, resulting in one physical neutral CP-even Higgs
boson, $\hsm$, after electroweak symmetry breaking (EWSB).  However,
the Standard Model is not likely to be the ultimate theoretical structure
responsible for electroweak symmetry breaking.
Moreover, the Standard Model must be viewed as an effective field theory
that is embedded in a more fundamental structure, characterized by an
energy scale, $\Lambda$, which is larger than the scale of EWSB,
$v=246\gev$.  Although $\Lambda$ may be as large as the Planck scale, 
there are strong theoretical arguments that suggest that $\Lambda$ is
significantly lower, perhaps of order 1 TeV~\cite{natural}. 
For example, $\Lambda$
could be the scale of supersymmetry 
breaking~\cite{susynatural,susyreview,susybreaking}, 
the compositeness scale
of new strong dynamics~\cite{techni}, 
or associated with the inverse size of extra
dimensions~\cite{extradim}. 
In many of these approaches, there exists an effective
low-energy theory with elementary scalars that comprise a
non-minimal Higgs sector~\cite{hhg}.  
For example, the minimal supersymmetric
extension of the Standard Model (MSSM) contains a scalar
Higgs sector corresponding to that of a two-Higgs-doublet model 
(2HDM)~\cite{susyhiggs,ghsusy}.
Models with Higgs doublets (and singlets) possess the important
phenomenological property that $\rho={\mw/(\mz \cos\theta_W)}=1$ up to
{\it finite} radiative corrections.

In this paper we focus on a general 2HDM.
There are two possible cases. In the first case, there is
never an energy range in which the effective low-energy
theory contains only one light Higgs boson. 
In the second case, one CP-even neutral Higgs boson,
$\hl$, is significantly lighter than a new scale, $\Lambda_{2HDM}$,
which characterizes the masses of all the remaining 2HDM Higgs states.
In this latter case, the scalar sector of the effective field theory below
$\lthdm$ is that of the SM Higgs sector.  In particular, if $\lthdm\gg v$, 
and all dimensionless Higgs self-coupling parameters
$\lam_i\lsim\mathcal{O}(1)$ [see \eq{pot}], then
the couplings of $\hl$ to gauge bosons and fermions and the
$\hl$ self-couplings approach the corresponding couplings of the
$\hsm$, with the deviations vanishing as some power of
$v^2/\lthdm^2$~\cite{habernir}.  
This limit is called the decoupling limit~\cite{decouplinglimit}, and is
one of the main subjects of this paper.

The purpose of this paper is to fully define and explore the
decoupling limit of the 2HDM.\footnote{Some of the topics of this
paper have also been addressed recently in \Ref{boudjema}.} 
We will explain the (often confusing)
relations between different parameter sets (\eg, Higgs masses and
mixing angles \vs\ Lagrangian tree-level couplings) and give a
complete translation table in Appendix~\ref{app:A}.  We then make one
simplifying assumption, namely that the Higgs sector is CP-conserving.
(The conditions that guarantee that there is no explicit or
spontaneous breaking of CP in the 2HDM are given in
Appendix~\ref{app:B}..  The
more general CP-violating 2HDM is treated
elsewhere~\cite{Dubinin:2002nx,kalinowski}.)  
In the CP-conserving 2HDM, there is
still some freedom in the choice of Higgs-fermion couplings.  A number
of different choices have been studied in the
literature~\cite{modeltypes,hhg}: type-I, in which only one Higgs
doublet couples to the fermions; and type-II, in which the neutral
member of one Higgs doublet couples only to up-type quarks and the
neutral member of the other Higgs doublet couples only to down-type
quarks and leptons.  For Higgs-fermion couplings of type-I or type-II,
tree-level flavor-changing neutral currents (FCNC) mediated by Higgs
bosons are automatically absent~\cite{higgsfcnc}.  
Type-I and type-II models can be
implemented with an appropriately chosen discrete symmetry (which may
be softly broken without dire phenomenologically consequences).  The
type-II model Higgs sector also arises in the MSSM.  In this paper, we
allow for the most general Higgs-fermion Yukawa couplings (the
so-called type-III model~\cite{typeiii}). For type-III Higgs-fermion
Yukawa couplings, tree-level Higgs-mediated FCNCs are present, and one
must be careful to choose Higgs parameters which ensure that these
FCNC effects are numerically small.  We will demonstrate in this paper
that in the approach to the decoupling limit, FCNC effects generated
by tree-level Higgs exchanges are suppressed by a factor of
$\mathcal{O}(v^2/\lthdm^2)$.

In Section~2, we define the most general CP-conserving 2HDM and
provide a number of useful relations among the parameters of the
scalar Higgs potential and the Higgs masses in Appendices~\ref{app:X}
and \ref{app:C}.  In Appendix~\ref{app:C2}, we note that certain
combinations of the scalar potential parameters are invariant with
respect to the choice of basis for the two scalar doublets.  In
particular, the Higgs masses and the physical Higgs interaction
vertices can be written in terms of these invariant coupling
parameters.  The decoupling limit of the 2HDM is defined in Section~3
and its main properties are examined.  In this limit, the properties
of the lightest CP-even Higgs boson, $\hl$, precisely coincide with
those of the SM Higgs boson.  This is shown in Section~4, where we
exhibit the tree-level Higgs couplings to vector bosons, fermions and
Higgs bosons, and evaluate them in the decoupling limit (cubic and
quartic Higgs self-couplings are written out explicitly in
Appendices~\ref{app:D} and \ref{app:E}, respectively).  The first
non-trivial corrections to the Higgs couplings as one moves away from
the decoupling limit are also given.  In Section~5, we note that
certain parameter regimes exist outside the decoupling regime in which
one of the CP-even Higgs bosons exhibits tree-level couplings that
approximately coincide with those of the SM Higgs boson.  We discuss
the origin of this behavior and show how one can distinguish this
region of parameter space from that of true decoupling.  In Section~6,
the two-Higgs-doublet sector of the MSSM is used to illustrate the
features of the decoupling limit when $\mha\gg\mz$.  In addition, we
briefly describe the impact of radiative corrections, and show how
these corrections satisfy the requirements of the decoupling limit.
We emphasize that the rate of approach to decoupling can be delayed at
large $\tan\beta$, and we discuss the possibility of a SM-like Higgs
boson in a parameter regime in which all Higgs masses are in a range
$\lsim\mathcal{O}(v)$.  Finally, our conclusions are give in
Section~7.

\section{The CP-Conserving Two-Higgs Doublet Model} \label{sec:two}      

We first review the general (non-supersymmetric)
two-Higgs doublet extension of the Standard Model \cite{hhg}.
Let $\Phi_1$ and
$\Phi_2$ denote two complex $Y=1$, SU(2)$\ls{L}$ doublet scalar fields.
The most general gauge invariant scalar potential is given
by\footnote{In \Refs{hhg}{ghsusy}, the scalar potential is parameterized
in terms of a different set of couplings, which are less
useful for the decoupling analysis.  In Appendix \ref{app:A}, we relate this
alternative set of couplings to the parameters appearing in \eq{pot}.}
\beqa
\mathcal{V}&=& m_{11}^2\Phi_1^\dagger\Phi_1+m_{22}^2\Phi_2^\dagger\Phi_2
-[m_{12}^2\Phi_1^\dagger\Phi_2+{\rm h.c.}]\nonumber\\[8pt]
&&\quad +\half\lambda_1(\Phi_1^\dagger\Phi_1)^2
+\half\lambda_2(\Phi_2^\dagger\Phi_2)^2
+\lambda_3(\Phi_1^\dagger\Phi_1)(\Phi_2^\dagger\Phi_2)
+\lambda_4(\Phi_1^\dagger\Phi_2)(\Phi_2^\dagger\Phi_1)
\nonumber\\[8pt]
&&\quad +\left\{\half\lambda_5(\Phi_1^\dagger\Phi_2)^2
+\big[\lambda_6(\Phi_1^\dagger\Phi_1)
+\lambda_7(\Phi_2^\dagger\Phi_2)\big]
\Phi_1^\dagger\Phi_2+{\rm h.c.}\right\}\,. \label{pot}
\eeqa
In general, $m_{12}^2$, $\lambda_5$,
$\lambda_6$ and $\lambda_7$ can be complex.
In many discussions of two-Higgs-doublet models, the terms proportional
to $\lambda_6$ and $\lambda_7$ are absent.  This can be achieved by
imposing a discrete symmetry $\Phi_1\to -\Phi_1$ on the model.  Such a
symmetry would also require $m_{12}^2=0$ unless we allow a
soft violation of this discrete symmetry by dimension-two 
terms.\footnote{This discrete 
symmetry is also employed to restrict the Higgs-fermion couplings
so that no tree-level Higgs-mediated FCNC's are present. If
$\lam_6=\lam_7=0$, but $m_{12}^2\neq 0$, the soft breaking
of the discrete symmetry generates {\it finite} Higgs-mediated
FCNC's at one loop.}
In this paper, we refrain in general from setting any of the coefficients
in eq.~(\ref{pot}) to zero.

We next derive the constraints on the parameters $\lam_i$ such that
the scalar potential $\mathcal{V}$ is bounded from below.  
It is sufficient to examine the quartic
terms of the scalar potential (which we denote by $\mathcal{V}_4$).
We define
$a\equiv\Phi_1^\dagger\Phi_1$, 
$b\equiv\Phi_2^\dagger\Phi_2$, 
$c\equiv {\rm Re}~\Phi_1^\dagger\Phi_2$, 
$d\equiv {\rm Im}~\Phi_1^\dagger\Phi_2$, and note that $ab\geq
c^2+d^2$.  Then, one can rewrite the quartic terms of the
scalar potential as follows:
\beqa
\mathcal{V}_4&=&\half\left[\lam_1^{1/2} a-\lam_2^{1/2} b\right]\lss{2}+
\left[\lam_3+(\lam_1\lam_2)^{1/2}\right](ab-c^2-d^2) \nonumber \\
&&+2[\lam_3+\lam_4+(\lam_1\lam_2)^{1/2}]\,c^2+[{\rm Re}~\lam_5-\lam_3
-\lam_4-(\lam_1\lam_2)^{1/2}](c^2-d^2) \nonumber \\[5pt]
&&-2cd\,{\rm Im}~\lam_5+2a\,[c\,{\rm Re}~\lam_6-d\,{\rm Im}~\lam_6]
+2b\,[c\,{\rm Re}~\lam_7-d\,{\rm Im}~\lam_7]\,.
\eeqa
We demand that no directions exist in field space in which
$\mathcal{V}\to -\infty$.  (We also require that no flat directions
exist for $\mathcal{V}_4$.)  
Three conditions on the $\lam_i$ are easily obtained by 
examining asymptotically large values of $a$ and/or $b$ with $c=d=0$:
\beq \label{globala}
\lam_1>0\,,\qquad \lam_2 >0\,,\qquad\lam_3>-(\lam_1\lam_2)^{1/2}\,.
\eeq
A fourth condition arises by examining the direction in field space
where $\lam_1^{1/2}a=\lam_2^{1/2}b$ and $ab=c^2+d^2$.  Setting $c=\xi d$,
and requiring that the potential is bounded from below for all $\xi$
leads to a condition on a quartic polynomial in $\xi$, which must be
satisfied for all $\xi$.  There is no simple analytical constraint on
the $\lam_i$ that can be derived from this condition.  If
$\lam_6=\lam_7=0$, the resulting polynomial is quadratic in $\xi$, and
a constraint on the remaining nonzero $\lam_i$ is easily
derived~\cite{bounded} 
\beq \label{globalb}
\lam_3+\lam_4-|\lam_5|>-(\lam_1\lam_2)^{1/2}\qquad [{\rm assuming}~
\lam_6=\lam_7=0]\,.
\eeq

In this paper, we
shall ignore the possibility of 
explicit CP-violating effects in the Higgs potential
by choosing all coefficients in eq.~(\ref{pot}) to be
real (see Appendix~B).\footnote{The most general 
CP-violating 2HDM will be examined in \Ref{kalinowski}.}
The scalar fields will
develop non-zero vacuum expectation values if the mass matrix
$m_{ij}^2$ has at least one negative eigenvalue.  
We assume that the
parameters of the scalar potential are chosen such that
the minimum of the scalar potential respects the
U(1)$\ls{\rm EM}$ gauge symmetry.  Then, the scalar field
vacuum expectations values are of the form
\beq
\langle \Phi_1 \rangle={1\over\sqrt{2}} \left(
\begin{array}{c} 0\\ v_1\end{array}\right), \qquad \langle
\Phi_2\rangle=
{1\over\sqrt{2}}\left(\begin{array}{c}0\\ v_2
\end{array}\right)\,,\label{potmin}
\eeq
where the $v_i$ are taken to be real, \ie\
we assume that spontaneous CP violation
does not occur.\footnote{The conditions required for the absence of
explicit and spontaneous CP-violation in the Higgs sector are
elucidated in Appendix \ref{app:B}.}
The corresponding potential minimum conditions are:
\beqa
m_{11}^2 &=& m_{12}^2\tb -\half
v^2\left[\lam_1\cb^2+\lamtil\sb^2
+3\lam_6\sb\cb+\lam_7\sb^2\tb\right]
\,,\label{minconditionsa} \\
m_{22}^2 &=& m_{12}^2\tb^{-1}-\half v^2
\left[\lam_2\sb^2+\lamtil\cb^2+\lam_6\cb^2\tb^{-1}+3\lam_7\sb\cb\right]\,,
\label{minconditionsb}
\eeqa
where we have defined:
\beq
\lamtil\equiv\lam_3+\lam_4+\lam_5\,,\qquad\qquad\tb\equiv\tanb\equiv{v_2\over
  v_1}\,,
\label{tanbdef}
\eeq
and
\beq
v^2\equiv v_1^2+v_2^2={4\mw^2\over g^2}=(246~{\rm GeV})^2\,.
\label{v246}
\eeq
It is always possible to choose the phases 
of the scalar doublet Higgs fields such that
both $v_1$ and $v_2$ are positive; henceforth we take
$0\leq\beta\leq\pi/2$.

Of the original eight scalar degrees of freedom, three Goldstone
bosons ($G^\pm$ and~$\go$) are absorbed (``eaten'') by the $W^\pm$ and
$Z$.  The remaining five physical Higgs particles are: two CP-even
scalars ($\hl$ and $\hh$, with $\mhl\leq \mhh$), one CP-odd scalar
($\ha$) and a charged Higgs pair ($\hpm$). 
The squared-mass parameters
$m_{11}^2$ and $m_{22}^2$ can be eliminated by minimizing the scalar
potential.  The resulting squared-masses for the CP-odd and charged
Higgs states are\footnote{Here and in the following, we use the
shorthand notation $\cb\equiv\cos\beta$, $\sb\equiv\sin\beta$,
$\ca\equiv\cos\alpha$, $\sa\equiv\sin\alpha$,
$\ctwoa\equiv\cos2\alpha$, $\stwoa\equiv\cos2\alpha$,
$\cbma\equiv\cos(\beta-\alpha)$, $\sbma\equiv\sin(\beta-\alpha)$, etc.}  
\beqa
\mha^2 &=&{m_{12}^2\over \sb\cb}-\half
v^2\big(2\lambda_5+\lambda_6\tb^{-1}+\lambda_7\tb\big)\,,\label{massha}\\[6pt]
m_{H^{\pm}}^2 &=& m_{A^0}^2+\half v^2(\lambda_5-\lambda_4)\,.
\label{mamthree}
\eeqa
\vskip1pc

\noindent
The two CP-even Higgs states mix according to the following squared-mass
matrix:
\beq
\calm^2 \equiv  m_{A^0}^2  \left(
   \begin{array}{cc}  \sb^2& -\sb\cb\\
                     -\sb\cb& \cb^2 \end{array}\right)
+\calb^2\,,
\label{massmhh}
\eeq
where
\beq
\calb^2\equiv v^2
\left( \begin{array}{cc}
  \lambda_1\cb^2+2\lambda_6\sb\cb+\lambda_5\sb^2
    &(\lambda_3+\lambda_4)\sb\cb+\lambda_6 \cb^2+\lambda_7\sb^2 \\[3pt]
 (\lambda_3+\lambda_4)\sb\cb+\lambda_6  \cb^2+\lambda_7\sb^2
    &\lambda_2\sb^2+2\lambda_7\sb\cb+\lambda_5\cb^2
\end{array}\right) \,.
\label{curlybdef}
\eeq
Defining the physical mass eigenstates
\beqa
\hh &=&(\sqrt{2}{\rm Re\,}\Phi_1^0-v_1)\ca+
(\sqrt{2}{\rm Re\,}\Phi_2^0-v_2)\sa\,,\nonumber\\
\hl &=&-(\sqrt{2}{\rm Re\,}\Phi_1^0-v_1)\sa+
(\sqrt{2}{\rm Re\,}\Phi_2^0-v_2)\ca\,,
\label{scalareigenstates}
\eeqa
the masses and mixing angle $\alpha$ are found from the diagonalization
process
\beqa
&& \!\!\!\!\!\!\!\!\!\!\!
\left(\begin{array}{cc} \mhh^2 & 0 \cr 0 &\mhl^2\end{array}\right)=
\left(\begin{array}{cc} \ca & \sa \cr -\sa & \ca \end{array}\right)
\left(\begin{array}{cc} \calm_{11}^2 & \calm_{12}^2 \\
\calm_{12}^2 &\calm_{22}^2\end{array}\right)
\left(\begin{array}{cc} \ca & -\sa \cr \sa & \ca \end{array}\right)
\nonumber\\[8pt]
&& \!\!\!
=\left(\begin{array}{cc} \calm_{11}^2\ca^2+2\calm_{12}^2\ca\sa
+\calm_{22}^2\sa^2 & \quad\calm_{12}^2(\ca^2-\sa^2)+(\calm_{22}^2-
\calm_{11}^2)\sa\ca \\[4pt]
\calm_{12}^2(\ca^2-\sa^2)+(\calm_{22}^2-\calm_{11}^2)\sa\ca
& \quad\calm_{11}^2\sa^2-2\calm_{12}^2\ca\sa
+\calm_{22}^2\ca^2 \end{array}\right).
\label{diagn}
\eeqa
The mixing angle $\alpha$ is evaluated by setting the off-diagonal
elements of the CP-even scalar squared-mass matrix [\eq{diagn}]
to zero, and demanding that $\mhh\geq\mhl$.  The end result is
\beq
 m^2_{\hh,\hl}=\half\left[{\cal M}_{11}^2+{\cal M}_{22}^2
\pm \sqrt{({\cal M}_{11}^2-{\cal M}_{22}^2)^2 +4({\cal M}_{12}^2)^2}
\ \right]\,.
\label{higgsmasses}
\eeq
and the corresponding CP-even scalar mixing angle is fixed by
\beqa
\stwoa &=&{2{\cal M}_{12}^2\over
\sqrt{({\cal M}_{11}^2-{\cal M}_{22}^2)^2 +4({\cal M}_{12}^2)^2}}\ ,\nonumber\\
\ctwoa &=&{{\cal M}_{11}^2-{\cal M}_{22}^2\over
\sqrt{({\cal M}_{11}^2-{\cal M}_{22}^2)^2 +4({\cal M}_{12}^2)^2}}\,.
\label{alphadef}
\eeqa
We shall take $-\pi/2\leq\alpha\leq\pi/2$.

It is convenient to define the following four combinations of parameters:
\beqa
m\ls{D}^4 &\equiv&\ \calb^2_{11}\calb^2_{22}-[\calb^2_{12}]^2\,,
\nonumber \\[3pt]
\mlsq &\equiv&\ \calb^2_{11}\cos^2\beta+\calb^2_{22}\sin^2\beta
+\calb^2_{12}\sin2\beta\,,\nonumber \label{mlsqdef} \\[3pt]
\mtsq &\equiv&\ \calb^2_{11}+\calb^2_{22}\,,\nonumber \\[3pt]
\mssq &\equiv&\ \mha^2+\mtsq\,,
\label{massdefs}
\eeqa
where the $\calb^2_{ij}$ are the elements of the matrix
defined in eq.~(\ref{curlybdef}).
In terms of these quantities we have the exact relations
\beq
 m^2_{\hh,\hl}=\half\left[\mssq\pm\sqrt{m\ls{S}^4-4\mha^2\mlsq
-4m\ls{D}^4}\,\right]\,,
\label{cpevenhiggsmasses}
\eeq
and
\beq
\cbma^2= {\mlsq-\mhl^2\over\mhh^2-\mhl^2}\,.
\label{cosbmasq}
\eeq
\Eq{cosbmasq} is most easily derived by using $\cbma^2={1\over 2}
(1+\ctwob\ctwoa+\stwob\stwoa)$ and the results of
eq.~(\ref{alphadef}).  Note that the case of $\mhl=\mhh$ is special
and must be treated carefully.  We do this in Appendix~\ref{app:X},
where we explicitly verify that $0\leq\cbma^2\leq 1$.

Finally, for completeness we record the expressions for the original
hypercharge-one scalar fields $\Phi_i$ in terms of the physical Higgs
states and the Goldstone bosons:
\beqa
\Phi_1^\pm &=& \cb G^\pm-\sb \hpm\,,\nonumber \\
\Phi_2^\pm &=& \sb G^\pm+\cb \hpm\,,\nonumber \\
\Phi_1^0 &=& \nicefrac{1}{\sqrt{2}}\left[v_1+\ca\hh-\sa\hl+i\cb
\go-i\sb\ha\right]\,,\nonumber \\ 
\Phi_2^0 &=& \nicefrac{1}{\sqrt{2}}\left[v_2+\sa\hh+\ca\hl+i\sb
\go+i\cb\ha\right]\,.
\label{heigenstates}
\eeqa

\section{The Decoupling Limit} \label{sec:twohalf}      

In effective field theory, we may examine the behavior of the theory
characterized by two disparate mass scales, $m_L\ll m_S$,  by
integrating out all particles with masses of order $m_S$,
assuming that all the couplings of the ``low-mass'' effective theory
comprising particles with masses of order $m_L$ can be kept fixed.
In the 2HDM, the low-mass effective theory, if it exists, must
correspond to the case where one of the Higgs doublets is integrated
out.  That is, the resulting effective low-mass theory is precisely
equivalent to the one-scalar-doublet SM Higgs sector.  These 
conclusions follow from electroweak gauge invariance.  Namely, there
are two relevant scales---the electroweak scale characterized by the
scale $v=246$~GeV and a second scale $m_S\gg v$.  The underlying
electroweak symmetry
requires that scalar mass splittings within doublets cannot be larger
than $\calo(v)$ [assuming that dimensionless couplings of the theory are
no larger than $\calo(1)$].  It follows that the $\hpm$, $\ha$ and $\hh$
masses must be of $\calo(m_S)$, while $\mhl\sim \calo(v)$.  Moreover,
since the effective low-mass theory consists of a one-doublet Higgs
sector, the properties of $\hl$ must be indistinguishable from those of
the SM Higgs boson.

We can illustrate these results more explicitly as follows.
Suppose that all the Higgs self-coupling constants $\lambda_i$ are
held fixed such that $|\lambda_i|\lsim \mathcal{O}(1)$, while taking
$\mha^2\gg|\lambda_i|v^2$.  In particular, we constrain
the $\alpha_i\equiv\lam_i/(4\pi)$ so that the Higgs sector does not become
strongly coupled, implying no violations of 
tree-unitarity \cite{clt,Lee:1977eg,weldon,arhrib,Kanemura:1993hm}.
Then, the $\calb^2_{ij}\sim{\cal O}(v^2)$, and
it follows that:
\beqa
\mhl &\simeq& m\ls{L}=\calo(v)\,,
\label{approxmasses1} \\[5pt]
\mhh,\mha,\mhpm &=&m_S+\calo\left(v^2/m_S\right)\,,
\label{approxmasses2}
\eeqa
\vspace{-0.2cm}
and
\vspace{-0.2cm}
\beqa
\cos^2(\beta-\alpha)&\simeq&\, {\mlsq(\mtsq-\mlsq)-m\ls{D}^4\over\mha^4}
\nonumber \\
&=&{\left[\half(\calb_{11}^2-\calb_{22}^2)\stwob-\calb_{12}^2\ctwob\right]^2
\over\mha^4}
=\calo\left({v^4\over m_S^4}\right)\,.
\label{approxcosbmasq}
\eeqa
We shall establish the above results in more detail below.

The limit $\mha^2\gg|\lambda_i|v^2$ (subject to $|\alpha_i|\lsim 1$) is 
called the {\it decoupling limit} of the model.\footnote{In 
Section~4 [see \eq{truedecoupling} and surrounding discussion],
we shall refine this definition slightly, and also require that
$\mha^2\gg |\lam_6|v^2\cotb$ and $\mha^2\gg |\lam_7|v^2\tanb$, in
order to guarantee that at large $\cotb$ [$\tanb$] the couplings of
$\hl$ to up-type [down-type] fermions approach the corresponding
SM Higgs-fermion couplings.\label{truedecoup}}
Note that eq.~(\ref{approxcosbmasq}) implies that 
in the decoupling limit, $\cbma=
{\cal O}(v^2/\mha^2)$. We will demonstrate
that this implies that the couplings of $\hl$ in the decoupling limit
approach values that correspond precisely to those of the SM Higgs boson. 
We will also obtain explicit expressions for
the squared-mass differences between the heavy Higgs bosons
(as a function of the $\lam_i$ couplings in the Higgs potential)
in the decoupling limit.

One can give an alternative condition for the decoupling limit.
As above, we assume that all $|\alpha_i|\lsim 1$.  First consider
the following special cases.
If neither $\tan\beta$ nor $\cot\beta$ is close to 0, then
$m_{12}^2\gg|\lambda_i|v^2$ [see \eq{massha}] in the decoupling limit.
On the other hand, if
$m_{12}^2\sim\calo(v^2)$ and $\tanb\gg 1$ [$\cot\beta \gg 1$], then it follows
from \eqns{minconditionsa}{minconditionsb} that $m_{11}^2\gg \calo(v^2)$ if
$\lambda_7< 0$
[$m_{22}^2\gg \calo(v^2)$ if $\lambda_6<0$] in the decoupling limit.
All such conditions
depend on the original choice of scalar field basis $\Phi_1$ and
$\Phi_2$.  For example, we can diagonalize the squared-mass terms of
the scalar potential [\eq{pot}] thereby setting $m_{12}=0$.  In the
decoupling limit in the new basis, one is simply driven to the second
case above.  A basis-independent
characterization of the decoupling limit is simple to formulate.
Starting from the scalar potential in an arbitrary basis,
form the matrix $m_{ij}^2$ [made up of the coefficients of the
quadratic terms in the potential, see \eq{pot}].
Denote the eigenvalues of this matrix by $m_a^2$ and $m_b^2$
respectively; note that the eigenvalues are real but can be of either
sign.  By convention, we can take $|m_a^2|\leq|m_b^2|$.  Then, the
decoupling limit corresponds to $m_a^2<0$, $m_b^2>0$ such that
$m_b^2\gg |m_a^2|, v^2$ (with $|\alpha_i|\lsim 1$).

For some choices of the scalar potential, no decoupling limit exists.
Consider the case of $m_{12}^2=\lam_6=\lam_7=0$ (and all other
$|\alpha_i|\lsim 1$).  Then, the potential minimum conditions
[\eqns{minconditionsa}{minconditionsb}]
do not permit either $m_{11}^2$ or $m_{22}^2$ to
become large;
$m_{11}^2$, $m_{22}^2\sim \calo(v^2)$, and clearly all Higgs masses are 
of $\calo(v)$.  Thus, in this case
no decoupling limit exists.\footnote{However, it may be
difficult to distinguish between the non-decoupling effects of the SM with a
heavy Higgs boson and those of the 2HDM where all Higgs bosons are
heavy~\cite{espriu}.}   
The case of
$m_{12}^2=\lam_6=\lam_7=0$ corresponds to the existence of a discrete
symmetry in which the potential is invariant under the change of sign
of one of the Higgs doublet fields.  Although the latter statement is
basis-dependent, one can check that the following stronger condition holds:
no decoupling limit exists if and only if $\lambda_6=\lambda_7=0$
in the basis where $m_{12}^2=0$.  Thus, the absence of a decoupling
limit implies the existence of some discrete symmetry under which the
scalar potential is invariant (although the precise form of this
symmetry is most evident for the special choice of basis).

We now return to the results for the Higgs masses and the CP-even Higgs
mixing angle in the decoupling limit.
For fixed values of $\lam_6$, $\lam_7$, $\alpha$ and $\beta$,
there are two
equivalent parameter sets: (i) $\lam_1$, $\lam_2$, $\lam_3$,
$\lam_4$ and $\lam_5$; (ii) $\mhl^2$, $\mhh^2$, $m_{12}^2$, $\mhpm^2$
and $\mha^2$. The relations between these two parameter sets are given
in Appendix \ref{app:C}. Using the results
\eqs{mhaform}{lamtomsqd} we can
give explicit expressions in the decoupling
limit for the Higgs masses in terms of the potential parameters
and the mixing angles.  First, it is convenient to define the following
four linear combinations of the $\lam_i$:\footnote{We make use of
 the triple-angle
identities: $\cthreeb=\cb(\cb^2-3\sb^2)$ and $\sthreeb=\sb(3\cb^2-\sb^2)$.}
\beqa
\!\!\!\!\!\!\!\!\!\!\!\!\lambar&\equiv& \lam_1\cb^4+\lam_2\sb^4
+\half\lamtil \stwob^2
+2\stwob(\lam_6\cb^2+\lam_7\sb^2)
\label{lambardef}
\,,   \\[5pt]
\!\!\!\!\!\!\!\!\!\!\!\!\lamhat&\equiv& \half\stwob\left[\lam_1\cb^2
-\lam_2\sb^2-\lamtil
  \ctwob\right]-\lam_6\cb\cthreeb-\lam_7\sb\sthreeb
\label{lamhatdef}
\,,\\[5pt]
\!\!\!\!\!\!\!\!\!\!\!\!\lam_{\ha}&\equiv&
\ctwob(\lam_1\cb^2-\lam_2\sb^2)+\lamtil s_{2\beta}^2-\lam_5
+2\lam_6\cb\sthreeb-2\lam_7\sb\cthreeb\,,\label{lamadef} \\[5pt]
\!\!\!\!\!\!\!\!\!\!\!\!\lamF&\equiv&\lam_5-\lam_4\,,
\label{lamfdef}
\eeqa
where $\lamtil$ is defined in \eq{tanbdef}.
The significance of these coupling combinations is discussed
in Appendix~\ref{app:C2}.
We consider the limit $\cbma\to 0$, corresponding to
the decoupling limit, $\mha^2\gg |\lam_i| v^2$.  
In nearly all of the 
parameter space, $\calm^2_{12}<0$ [see \eq{massmhh}], and it follows from
\eq{alphadef} that $-\pi/2\leq\alpha\leq 0$ (which implies that
$\cbma\to 0$ is equivalent to
$\beta-\alpha\to\pi/2$ given that $0\leq\beta\leq\pi/2$).
However, in the small regions of parameter space in which $\beta$ is
near zero [or $\pi/2$], 
roughly corresponding to $\mha^2\tanb<\lambda_6 v^2$
[or $\mha^2\cot\beta<\lambda_7 v^2$],
one finds $\calm^2_{12}>0$ (and consequently
$0<\alpha<\pi/2$).  In these last two cases, the decoupling limit
is achieved for $\alpha=\pi/2-\beta$ and $\cot\beta\gg 1$ [$\tanb\gg 1$].
That is, $\cos(\beta-\alpha)=\sin 2\beta\ll 1$ and 
$\sin(\beta-\alpha)\simeq -1$ [$+1$].~\footnote{We have chosen
a convention in which $-\pi/2\leq\alpha\leq\pi/2$. An equally good
alternative is to choose $\sin(\beta-\alpha)\geq 0$. If negative,
one may simply change the sign of $\sin(\beta-\alpha)$ by taking
$\alpha\to \alpha \pm \pi$, which is equivalent to the field redefinitions
$\hl\to -\hl$, $\hh\to -\hh$.}
In practice, since $\tan\beta$ is fixed and 
cannot be arbitrarily large (or arbitrarily close to zero), one can always 
find a value of $\mha$ large enough such that $\calm^2_{12}<0$.
This is equivalent to employing the refined version of the decoupling
limit mentioned in footnote\fnref{truedecoup}.  In this case,
the decoupling limit simply corresponds to 
$\beta-\alpha\to \pi/2$ 
[{\it i.e.}, $\sin(\beta-\alpha)=1$] independently of the value of $\beta$.  

In the approach to the decoupling limit where $\alpha\simeq\beta-\pi/2$
(that is, $|\cbma|\ll 1$ and $\sbma\simeq 1-\half\cbma^2$), we may use
\eqs{altmha}{mhhmhadiff} and \eq{mamthree} to obtain:\footnote{In obtaining
eqs.~(\ref{decouplimits1}), (\ref{decouplimits3}) and (\ref{decouplimits4})
we divided both sides of each equation by $\cbma$, so these equations
need to be treated with care if $\cbma=0$ exactly.  In this latter
case, it suffices to note that $\lamhat/\cbma$ has a finite limit
whose value depends on $\mha$ and $\lamA$ [see \eq{cbmadec}].}
\beqa
\mha^2 &\simeq &
v^2 \left[{\lamhat\over\cbma}+\lam_A-\nicefrac{3}{2}\lamhat\,\cbma
\right]\,,\label{decouplimits1}\\
\mhl^2 &\simeq & v^2(\lambar-\lamhat\,\cbma)\,,\label{decouplimits2}\\
\mhh^2 &\simeq &
v^2 \left[{\lamhat\over\cbma}+\lam-\half\lamhat\,\cbma
\right]\simeq \mha^2
+(\lambda-\lambda_A+\lamhat\,\cbma)v^2\,,\label{decouplimits3}\\
\mhpm^2 &\simeq &
v^2 \left[{\lamhat\over\cbma}+\lam_A+\half\lamF
-\nicefrac{3}{2}\lamhat\,\cbma\right]=\mha^2+\half\lamF v^2\,.
\label{decouplimits4}
\eeqa
The condition $\mhh>\mhl$ implies
the inequality (valid to first order in $\cbma$):
\beq
\mha^2> v^2(\lam_A-2\lamhat\cbma)\,,
\label{mhacond}
\eeq
[{\it cf.}~\eq{mhaineq}].
The positivity of $\mhl^2$ also imposes a useful constraint on
the Higgs potential parameters.  For example, 
$\mhl^2> 0$ requires that $\lambar>0$.

In the decoupling limit (where $\mha^2\gg|\lam_i|v^2$),
\eqs{decouplimits1}{decouplimits4} provide the first nontrivial corrections to
\eqns{approxmasses1}{approxmasses2}.
Finally, we employ \eq{massha} to obtain
\beq
m_{12}^2 \simeq
v^2\sb\cb \left[{\lamhat\over\cbma}+
\lam_{\ha}+\lam_5+\half\lam_6\tb^{-1}+\half\lam_7\tb-
\nicefrac{3}{2}\lamhat\cbma\right]\,.
\eeq
This result confirms our previous observation that $m_{12}^2\gg
|\lambda_i| v^2$ in the decoupling limit
as long as $\beta$ is not close to $0$ or $\pi/2$.
However, $m_{12}^2$ can be of ${\mathcal{O}}(v^2)$ in the decoupling
limit [$\cbma\to 0$] 
if either $t_\beta\gg 1$ [and $\cb/\cbma\sim {\mathcal{O}}(1)$]
or $\tb^{-1}\gg 1$ [and $\sb/\cbma\sim {\mathcal{O}}(1)$].

The significance of \eq{decouplimits2}
is easily understood by noting that the decoupling limit corresponds to
integrating out the second heavy Higgs doublet.  The resulting low-mass
effective theory is simply the one-Higgs-doublet model with
corresponding scalar potential
$V=m^2(\Phi^\dagger\Phi)+\half\lambar(\Phi^\dagger\Phi)^2$, where
$\lambar$ is given by \eq{lambardef} and 
\beq
m^2\equiv m_{11}^2\cb^2+m_{22}^2\sb^2-2m_{12}^2\sb\cb\,.
\eeq
Imposing the potential minimum
conditions [\eqns{minconditionsa}{minconditionsb}], we see that 
$v^2=-2m^2/\lambar$ [where $\vev{\Phi^0}\equiv v/\sqrt{2}$] 
as expected.  Moreover, the Higgs mass
is given by $\mhl^2= \lambar v^2$, in agreement with the $\cbma\to
0$ limit of \eq{decouplimits2}.

We can rewrite \eq{decouplimits1} in another form
[or equivalently use \eqns{exactbma1}{exactbma2} to obtain]:
\beq
\cos(\beta-\alpha)\simeq  {\lamhat v^2 \over \mha^2-\lambda_A v^2}
\label{cbmadec} 
\simeq{\lamhat v^2\over \mhh^2-\mhl^2}\,.
\eeq
This yields an ${\mathcal O}(v^2/\mha^2)$ correction   
to \eq{approxcosbmasq}.
Note that \eq{cbmadec} also implies that in the approach to the decoupling
limit, the sign of
$\cos(\beta-\alpha)$ is given by the sign of $\lamhat$.

\section{Two-Higgs Doublet Model Couplings in the Decoupling Limit}
\label{sec:three}

The phenomenology of the two-Higgs doublet model depends in detail on
the various couplings of the Higgs bosons to gauge bosons, Higgs
bosons and fermions~\cite{hhg}.  
The Higgs couplings to gauge bosons follow from
gauge invariance and are thus model independent:
\beq
 g\ls{\hl VV}=g\ls{V} m\ls{V}\sbma\,,\quad
           g\ls{\hh VV}=g\ls{V} m\ls{V}\cbma\,,\label{vvcoup}
\eeq
where $g_V\equiv 2m_V/v$ for $V=W$ or $Z$.
There are no tree-level couplings of $\ha$ or $\hpm$ to $VV$.
In the decoupling limit where $\cbma=0$, we see that $g_{\hl
VV}=g_{\hsm VV}$, whereas the $\hh VV$ coupling vanishes. 
Gauge invariance also determines the strength of the trilinear
couplings of one gauge boson to two Higgs bosons:
\beq g\ls{\hl\ha Z}={g\cbma\over 2\cos\theta_W}\,,\quad
           g\ls{\hh\ha Z}={-g\sbma\over 2\cos\theta_W}\,.
           \label{hvcoup}
\eeq
In the decoupling limit, the $\hl\ha Z$ coupling vanishes, while
the $\hh\ha Z$ coupling attains its maximal value.
This pattern is repeated in all the three-point and four-point
couplings of $\hl$ and $\hh$ to $VV$, $V\phi$, and $VV\phi$ final
states (where $V$ is a vector boson and $\phi$ is one of the Higgs
scalars).  These results can be summarized as follows:
the coupling of $\hl$ and $\hh$ to
vector boson pairs or vector--scalar boson final states is proportional
to either $\sin(\beta-\alpha)$ or $\cos(\beta-\alpha)$ as indicated
below \cite{hhg,ghsusy}.
\beq
\renewcommand{\arraycolsep}{2cm}
\let\us=\underline
\begin{array}{ll}
  \us{\cos(\beta-\alpha)}&  \us{\sin(\beta-\alpha)}\\[3pt]
       \hh W^+W^-&        \hl W^+W^- \\
       \hh ZZ&            \hl ZZ \\
       Z\ha\hl&          Z\ha\hh \\
       W^\pm H^\mp\hl&  W^\pm H^\mp\hh \\
       ZW^\pm H^\mp\hl&  ZW^\pm H^\mp\hh \\
       \gamma W^\pm H^\mp\hl&  \gamma W^\pm H^\mp\hh
\end{array}
\label{littletable}
\eeq
Note in particular that {\it all} vertices
in the theory that contain at least
one vector boson and {\it exactly one} of the non-minimal Higgs boson states
($\hh$, $\ha$ or $\hpm$) are proportional to $\cos(\beta-\alpha)$ and
hence vanish in the decoupling limit.

The Higgs couplings to fermions are model dependent.
The most general structure for the Higgs-fermion Yukawa couplings,
often referred to as the type-III model \cite{typeiii}, is given by:
\beq \label{ymodeliii}
\!\!\!\!\!\!\!\!
-\call_Y=\anti \qlo \wtil\Phi_1\eiuo  \uro +\anti Q_L^0\Phi_1\eido \dro
+ \anti \qlo \wtil\Phi_2\eiiuo \uro +\anti \qlo \Phi_2\eiido \dro
+{\rm h.c.}\,,
\eeq
where $\Phi_{1,2}$ are the Higgs doublets, $\wtil\Phi_i\equiv
i\sigma_2 \Phi^*_i$,
$\qlo $ is the weak isospin quark doublet, 
and $\uro$, $\dro$ are weak isospin quark singlets.
[The right and left-handed fermion fields are defined as usual:
$\psi_{R,L}\equiv P_{R,L}\psi$, where $P_{R,L}\equiv \half(1\pm\gamma_5)$.]
Here, $\qlo $, $\uro $,
$\dro $ denote the interaction basis states, which
are vectors in flavor space, whereas $\eiuo,\eiiuo,\eido,\eiido$ are matrices
in flavor space.  We have omitted the leptonic couplings in \eq{ymodeliii}; 
these follow the same pattern as the down-type quark couplings.

We next shift the scalar fields according to their vacuum expectation
values, and then re-express the scalars in terms of the physical Higgs
states and Goldstone bosons [see \eq{heigenstates}].  
In addition, we diagonalize the quark
mass matrices and define the quark mass eigenstates.
The resulting Higgs-fermion Lagrangian can be written in
several ways \cite{Diaz:dh}.  We choose to display the form that
makes the type-II model limit 
of the general type-III couplings apparent.
The type-II model (where $\eiuo=\eiido=0$) automatically has no tree-level
flavor-changing neutral Higgs couplings, whereas
these are generally present for type-III couplings.
The fermion mass eigenstates are related to the interaction eigenstates
by biunitary transformations:
\beqa
&& P_L U=V_L^U P_L U^0\,,\qquad P_R U=V_R^U P_R U^0\,,\nonumber \\
&& P_L D=V_L^D P_L D^0\,,\qquad P_R D=V_R^D P_R D^0\,,
\eeqa
and the Cabibbo-Kobayashi-Maskawa matrix is defined as $K\equiv V_L^U
V_L^{D\,\dagger}$.  It is also convenient to define ``rotated''
coupling matrices:
\beq \label{etayuks}
\eiui\equiv V_L^U \eiuoi V_R^{U\,\dagger}\,,\qquad
\eidi\equiv V_L^D \eidoi V_R^{D\,\dagger}\,.
\eeq
The diagonal quark mass matrices are obtained by replacing the scalar
fields with their vacuum expectation values:
\beq
\mdd={1\over \sqrt 2}(v_1\eid+v_2\eiid)\,,\quad
\mud={1\over \sqrt 2}(v_1\eiu+v_2\eiiu)\,.
\eeq
After eliminating $\eiiu$ and $\eid$, the 
resulting Yukawa couplings are:
\beqa
\call_Y&=&
{1\over v}\,\anti D\mdd D\left({\sa\over \cb}\hl-{\ca\over \cb}\hh\right)
+{i\over v}\,\anti D\mdd\gamma_5 D(\tb\ha-\go)\nonumber \\
&&-{1\over \rtwo\cb}\anti D (\eiid P_R+ {\eiid}^\dagger P_L)D
(\cbma\hl-\sbma\hh)
-{i\over \rtwo\cb}\anti D (\eiid P_R- {\eiid}^\dagger P_L)D\,\ha
\nonumber\\
&&-{1\over v}\,\anti U\mud U\left({\ca\over \sb}\hl+{\sa\over \sb}\hh\right)
+{i\over v}\,\anti U\mud\gamma_5 U(\tbi\ha+\go)\nonumber\\
&&+{1\over \rtwo\sb}\anti U (\eiu P_R+ {\eiu}^\dagger P_L)U 
(\cbma\hl-\sbma\hh)
-{i\over \rtwo\sb}\anti U (\eiu P_R- {\eiu}^\dagger P_L)U\,\ha 
\nonumber\\
&&+{\rtwo\over v}\biggl[\anti U K\mdd P_R D(\tb\hp-G^+)
+\anti U \mud K P_LD(\tbi\hp+G^+)+{\rm h.c.}\biggr]\nonumber\\
&&-\left[{1\over \sb}\anti U {\eiu}^\dagger K P_LD\,\hp
+{1\over \cb}\anti U K\eiid  P_RD\,\hp +{\rm h.c.}\right]\,.
\label{yukform}
\eeqa
In general, $\eiu$ and $\eiid$ are complex non-diagonal matrices.
Thus, the Yukawa Lagrangian displayed in \eq{yukform} exhibits both
flavor-nondiagonal and CP-violating couplings between the neutral
Higgs bosons and the quarks.  

In the decoupling limit (where $\cbma\to 0$), the Yukawa Lagrangian
displays a number of interesting features.  First, the flavor
non-diagonal and the CP-violating couplings of $\hl$ vanish (although
the corresponding couplings to $\hh$ and $\ha$ persist). Moreover,
in this limit, the $\hl$ coupling to fermions reduces precisely to its
Standard Model value, $\call_Y^{\rm SM}=-(m_f/v)\bar ff\hl$.
To better see the behavior of couplings in the decoupling limit, the
following trigonometric identities are particularly useful:
\beqa
\hl\overline D D&:&~~ -{\sin\alpha\over\cos\beta}=\sin(\beta-\alpha)
-\tan\beta\cos(\beta-\alpha)\,,\label{qqcouplings1}\\[3pt]
\hl\overline U U&:&~~~ \phm{\cos\alpha\over\sin\beta}=\sin(\beta-\alpha)
+\cot\beta\cos(\beta-\alpha)\,,\label{qqcouplings2}\\[3pt]
\hh\overline D D&:&~~~ \phm{\cos\alpha\over\cos\beta}=\cos(\beta-\alpha)
+\tan\beta\sin(\beta-\alpha)\,,\label{qqcouplings3}\\[3pt]
\hh \overline U U&:&~~~ \phm{\sin\alpha\over\sin\beta}=\cos(\beta-\alpha)
-\cot\beta\sin(\beta-\alpha)\label{qqcouplings4}\,,
\eeqa
where we have indicated the type of Higgs-fermion coupling with which
a particular trigonometric expression arises.  It is now easy to read
off the corresponding Higgs-fermion couplings in the decoupling limit
and one verifies that the $\hl$-fermion couplings reduce to their
Standard Model values.  Working to $\mathcal{O}(\cbma)$, the Yukawa
couplings of $\hl$ are given by
\beqa \label{hlqqlag}
\!\!\!\!\!\!\!\!\!\!
\call_{\hl QQ} =  &\,- \anti D\, & \left[{1\over v}M_D-\tanb
\left[{1\over v}M_D
-{1\over\sqrt{2}\sb}\left(S_D+iP_D\gamma_5\right)\right]\cbma\right]D
\,\hl\nonumber
\\
 &\,- \anti U\, & \left[{1\over v}M_U+\cotb\left[{1\over v}M_U
-{1\over\sqrt{2}\cb}\left(S_U+iP_U\gamma_5\right)\right]\cbma\right]U
\,\hl\,,
\eeqa
where 
\beq \label{sdpd}
S_D  \equiv \half\left(\eta_2^D+\eta_2^{D\,\dagger}\right)\,,\qquad\qquad
P_D  \equiv -\nicefrac{i}{2}\left(\eta_2^D-\eta_2^{D\,\dagger}\right)\,,
\eeq
are $3\times 3$ hermitian matrices and $S_U$ and $P_U$ are defined
similarly by making the replacements $D\to U$ and $2\to 1$.  Note that
both $\hl$-mediated FCNC interactions (implicit in the
off-diagonal matrix elements of $S$ and $P$) and CP-violating
interactions proportional to $P$ are suppressed by a factor of $\cbma$  
in the decoupling limit.  
Moreover, FCNCs and CP-violating effects mediated by 
$\ha$ and $\hh$ are suppressed by the square of the
heavy Higgs masses (relative to $v$), due to the propagator suppression.
Since $\mhl\ll\mhh$, $\mha$ and 
$\cbma\simeq \calo(v^2/\mha^2)$ near the decoupling limit, we see that the
flavor-violating processes and CP-violating processes
mediated by $\hl$, $\hh$ and $\ha$ are all
suppressed by the same factor.
Thus, for $\mha\gsim\mathcal{O}(1~{\rm TeV})$,
the decoupling limit provides a viable mechanism 
for suppressed
Higgs-mediated FCNCs and suppressed 
Higgs-mediated CP-violating effects in the most general 2HDM.

Note that the approach to decoupling can be 
delayed if either $\tanb\gg 1$ or $\cot\beta\gg 1$, as is evident 
from \eq{hlqqlag}.  For example, decoupling 
at large $\tanb$ or $\cot\beta$ occurs when 
$|\cbma\tanb|\ll 1$ or $|\cbma\cot\beta|\ll 1$, respectively.  Using 
\eqns{cbmadec}{lamhatdef}, these conditions are respectively
equivalent to 
\beq \label{truedecoupling}
\mha^2\gg |\lam_6|v^2\cot\beta \qquad {\rm and} \qquad
\mha^2\gg |\lam_7|v^2\tan\beta\,,
\eeq
which supplement the usual requirement of $\mha^2\gg\lam_i v^2$.
That is, there are two possible ranges of the CP-odd Higgs 
squared-mass, $\lam_i
v^2\ll\mha^2\lsim |\lam_7|v^2\tanb$ [or $\lam_i
v^2\ll\mha^2\lsim|\lam_6|v^2\cot\beta$] when $\tanb\gg 1$
[or $\cot\beta\gg 1$], where the $\hl$ couplings to $VV$, $\hl\hl$ and
$\hl\hl\hl$ are nearly indistinguishable
from the corresponding $\hsm$ couplings, whereas one of the $\hl f\bar f$ 
couplings can deviate significantly 
from the corresponding $\hsm f\bar f$ couplings.

The cubic and quartic Higgs self-couplings depend on the
parameters of the 2HDM potential [eq.~(\ref{pot})], and
are listed in Appendices~\ref{app:D} and \ref{app:E}, respectively. 
In the decoupling limit (DL) of $\alpha\to\beta-\pi/2$,
we denote the terms of the scalar potential corresponding to the
cubic Higgs couplings by $\mathcal{V}_{\rm DL}^{(3)}$ and 
the terms corresponding to the quartic
Higgs couplings by $\mathcal{V}_{\rm DL}^{(4)}$.
The coefficients of the
quartic terms in the scalar Higgs potential 
can be written more simply in terms of the
linear combinations of couplings defined earlier [\eqs{lambardef}{lamfdef}]
and three additional combinations (see Appendix~\ref{app:C2} for a
discussion of the significance of these combinations):
\beqa
\lamT &=&
\quarter\stwob^2(\lam_1+\lam_2)+\lamtil(\sb^4+\cb^4)-2\lam_5-s_{2\beta}
c_{2\beta}(\lam_6-\lam_7)\,, \label{lamtdef}\\[5pt]
\lamU &=&\half s_{2\beta}(\sb^2\lam_1-\cb^2\lam_2+
c_{2\beta}\lamtil)-\lam_6\sb\sthreeb -\lam_7\cb\cthreeb\,.
\label{lamudef}\\[5pt] 
\lamV &=& \lam_1\sb^4+\lam_2\cb^4+\half\lamtil s_{2\beta}^2-
2s_{2\beta}(\lam_6\sb^2+\lam_7\cb^2)\,. 
\label{lamvdef}
\eeqa
The resulting expressions for  $\mathcal{V}_{\rm DL}^{(3)}$ and   
$\mathcal{V}_{\rm DL}^{(4)}$ are
\beqa  \label{vcubic}
{\cal V}_{\rm DL}^{(3)}&=&
\half\lam v(\hl^3+\hl\go^2+2\hl\gp\gm) 
+(\lamT+\lamF)v \hl\hp\hm\nncr &&
+\half\lamhat v\left[3\hh\hl^2+\hh\go^2+2\hh\gp\gm -2
\hl(\ha\go+\hp\gm+\hm\gp)\right] \nncr &&
+\half\lamU v (\hh^3+\hh\ha^2+2\hh\hp\hm)\nncr &&
+\left[\lamA-\lam+\half\lamF\right] v \hh(\hp\gm+\hm\gp) \nncr &&
+(\lamA-\lam) v \hh\ha\go +\half\lamT v \hl \ha^2 
+(\lam-\lamA+\half\lamT) v \hl\hh^2 \nncr &&
+\nicefrac{i}{2}\lamF v \ha (\hp\gm-\hm\gp)\,,
\eeqa
and
\beqa
{\cal V}_{\rm DL}^{(4)}&&=
\eighth\lam (\g^2+2 \gp \gm+\hl^2)^2 
\nncr
&&+\lamhat(\hl^3 \hh-\hl^2 \ha \g-\hl^2 \hp \gm-\hl^2 \hm
\gp+\hl \hh \g^2+2 \hl \hh \gp \gm-\ha \g^3
\nncrsmall
&&\quad-2 \ha \g \gp \gm-
\g^2 \hm \gp-\g^2 \hp \gm-2 \hp \gm\gp \gm-2 \hm \gp \gm\gp)
\nncr
&&+\half(\lamT+\lamF)
(\hl^2 \hp \hm+\hh^2 \gp \gm+\ha^2 \gp \gm+\g^2 \hp \hm) 
\nncr
&&+\lamU
(\hl \hh^3+\hl \hh \ha^2+2 \hl \hh \hp \hm
-\hh^2 \ha \g-\hh^2 \hp \gm-\hh^2 \hm \gp-\ha^3 \g
\nncrsmall
&&\quad
-\ha^2 \hp \gm-\ha^2 \hm \gp-2 \ha \g \hp \hm
-2 \hp \hm \hp \gm-2 \hm\hp \hm \gp)
\nncr
&&+\left[2(\lamA-\lam)+\lamF\right](\hl \hh \hp \gm+\hl \hh \hm \gp
-\ha \g \hp \gm-\ha\g \hm \gp) 
\nncr
&&+\quarter\lamV(\hh^4+2 \hh^2 \ha^2+\ha^4+4 \hh^2 \hp \hm+4 \ha^2 \hp \hm+
4 \hp\hm\hp \hm) 
\nncr
&&+\half(\lam-\lamA)(\hp\hp \gm\gm+\hm\hm \gp\gp-2 \hl \hh \ha \g) 
+\quarter\lamT(\hl^2 \ha^2+\hh^2 \g^2) 
\nncr&&
+\quarter\left[2(\lam-\lamA)+\lamT\right](\hl^2 \hh^2+\ha^2 \g^2) 
+(\lam-\lamA+\lamT)\hp \hm \gp \gm 
\nncr
&&+\nicefrac{i}{2}\lamF
(\hl \ha \hp \gm-\hl \ha \hm \gp+\hh \g \hp \gm
-\hh \g \hm \gp)\,,
\label{vquartic}
\eeqa
where $\go$ and $\gpm$ are the Goldstone bosons (eaten
by the $Z$ and $\wpm$, respectively).
Moreover, for $\cbma=0$, we have $\mhl^2 = \lambda v^2$ and
$\mhh^2-\mha^2 = (\lambda-\lamA) v^2$,
whereas $\mhpm^2-\mha^2 = \half\lamF v^2$ is exact at tree-level.
As expected, in the decoupling limit, the low-energy effective scalar theory
(which includes $\hl$ and the three Goldstone bosons) is precisely the
same as the corresponding SM Higgs theory, with $\lambda$ proportional
to the Higgs quartic coupling.

One can use the results of Appendices~\ref{app:D} and \ref{app:E}
to compute the first non-trivial $\mathcal{O}(\cbma)$
corrections to \eqns{vcubic}{vquartic} as one moves away
from the decoupling limit.
These results are given in
Tables~\ref{hhhdecouptab} and~\ref{hhhhdecouptab}.  For example, the
$\hl\hl\hl$ and $\hl\hl\hl\hl$ couplings in the decoupling limit are given by 
\beqa
g_{\hl\hl\hl}\,&\simeq\, & -3v(\lam-3\lamhat\cbma)\simeq
{- 3\mhl^2\over v}+6\lamhat\cbma v
\,, 
\label{hhhdecoup}\\
g_{\hl\hl\hl\hl}\,&\simeq\, &  -3(\lam-4\lamhat\cbma)\simeq
{- 3\mhl^2\over v^2}+9\lamhat\cbma
\,,
\label{hhhhdecoup}
\eeqa
where we have used \eq{decouplimits2}.
Precision measurements of these couplings could 
in principle (modulo radiative corrections, 
which are known within the SM~\cite{sirlin}) provide 
evidence for a departure from the corresponding SM relations.

Using the explicit forms for the quartic Higgs
couplings given in Appendix~\ref{app:E}, it follows that 
all quartic couplings are $\lsim \mathcal{O}(1)$
if we require that the $\lam_i\lsim \mathcal{O}(1)$.  
Unitarity constraints on
Goldstone/Higgs scattering processes 
can be used to impose numerical limitations on
the contributing quartic 
couplings~\cite{clt,Lee:1977eg,weldon,arhrib,Kanemura:1993hm}.  
If we apply tree-level unitarity
constraints for $\sqrt{s}$ larger than all Higgs masses, then 
${\lam_i/ 4\pi}\lsim\mathcal{O}(1)$ (the precise analytic
upper bounds are given in \Ref{arhrib}).   One can also investigate a
less stringent requirement if the Higgs sector is close to the
decoupling limit.  Namely, assuming $\mhl\ll \mhh,\mha,\mhpm$, one can
simply impose unitarity constraints on the low-energy effective scalar
theory.   One must check, for example, that all $2\to
2$ scattering processes involving the $W^\pm$, $Z$ and $\hl$ 
satisfy partial-wave unitarity~\cite{Lee:1977eg,arhrib,Kanemura:1993hm}.  
At tree-level, one simply obtains the well
known SM result, $\lam\leq 8\pi/3$, where $\lam$ is 
given by \eq{lambardef}.\footnote{
Using $\mhl^2=\lam v^2$, this bound is a factor of 2
more stringent than that of \Ref{Lee:1977eg} based on the
requirement $|{\rm Re}~a_0|\leq \half$ for the $s$-wave partial wave
amplitude \cite{Luscher:1988gc}.}
At one-loop, the heavier Higgs scalars can contribute via virtual
exchanges, and the restrictions on the self-couplings
now involve both the light and the heavier Higgs scalars. For example,
in order to avoid large one-loop corrections to the
four-point interaction $W^+W^-\to \hl\hl$ via an intermediate loop of
a heavy Higgs pair, the quartic
interactions among $\hl^2\hh^2$, $\hl^2\ha^2$ and $\hl^2\hp\hm$ must be
perturbative.   In this
case, \eq{vquartic} implies that   $|\lam-\lam_A|,~|\lamF|\lsim 1$.
It follows that there is a bound on 
the squared-mass splittings among the heavy Higgs states
of $\calo(v^2)$.
Thus, to maintain unitarity and perturbativity,
the decoupling limit demands rather degenerate heavy Higgs bosons.

Using the explicit forms for the cubic Higgs
couplings given in Appendix~\ref{app:D}, it follows that 
all cubic couplings are $\lsim \mathcal{O}(v)$
if we require that the $\lam_i\lsim \mathcal{O}(1)$.
The cubic couplings can also be rewritten in terms of the Higgs masses. 
For example, one possible form for the $\hl\hl\hl$ coupling is given
in \eq{hhhalt}.  Here, we shall consider two equivalent expressions
for the $\hl H^+H^-$ coupling:  
\beqa \label{gcubic}
&&g\ls{\hl\hp\hm}=-{1\over v}\left[\left(\mhl^2-{m_{12}^2\over \sb\cb}\right)
{\cbpa\over\sb\cb}+\left(2\mhpm^2-\mhl^2\right)\sbma
+\half v^2\left({\lam_6\over\sb^2}-{\lam_7\over\cb^2}\right)\cbma
\right] \nonumber \\[8pt]
&&\quad ={1\over v}\left[(2\mha^2-2\mhpm^2-\mhl^2)\sbma+2(\mha^2-\mhl^2)
{c_{2\beta}\cbma\over s_{2\beta}}
+v^2\left({\lam_5\cbpa\over\sb\cb}-{\lam_6\sa\over\sb}+{\lam_7\ca\over\cb}
\right)\right] \,.\nonumber \\[8pt]
&&\phantom{equation}
\eeqa
From the first line of
\eq{gcubic}, it appears that $g\ls{\hl\hp\hm}$ grows 
quadratically with the heavy charged Higgs mass.  However, this is an 
illusion, as can be seen in the subsequent expression for
$g\ls{\hl\hp\hm}$.  In particular, 
$\mha^2-\mhpm^2\sim\mathcal{O}(v^2)$ follows from 
\eq{mamthree}, while in the decoupling limit, 
$\mha^2\cbma\sim\mathcal{O}(v^2)$ follows from 
\eq{mhaform}.  Hence,
$g\ls{\hl\hp\hm}\sim\mathcal{O}(v)$ as expected.  
One can also check that the apparent 
singular behavior as $\sb\to 0$ or $\cb\to 0$ is in fact absent,
since the original form of $g\ls{\hl\hp\hm}$ was well behaved in this limit.
Clearly,
the most elegant form for $g_{\hl\hp\hm}$ is given in \eq{invdefghaa}.
No matter which form is used, it is straightforward to perform an expansion 
for small $\cbma$ to obtain 
\beq
g\ls{\hl\hp\hm}=-v(\lamF+\lam_T)+\mathcal{O}(\cbma)\,,
\eeq
which agrees with the corresponding result given in
Table~\ref{hhhdecouptab} of Appendix~\ref{app:D}.

One can also be misled by writing the cubic couplings in terms of
$\Lambda_i$, which are employed in an alternate parameterization of the
2HDM scalar potential given in Appendix~\ref{app:A}.  In particular, in the 
CP-conserving case, $m_{12}^2=\half v^2\sb\cb\Lambda_5$, 
which becomes large in the approach to the decoupling limit.  
Consequently, all the $\Lambda_i$ ($i=1,\ldots,6$)
are large in the decoupling limit
[see \eq{convertback}], even though the magnitudes of the  
$\lam_i$ are all $\lsim\mathcal{O}(1)$.  

One important consequence of $g\ls{\hl\hp\hm}\sim\mathcal{O}(v)$ is
that the one-loop amplitude for $\hl\to\gamma\gamma$ reduces to the
corresponding SM result in the decoupling limit
(where $\mhpm\gg v$).  To prove this, we observe that
in the decoupling limit, all $\hl$ couplings to SM particles
that enter the one-loop Feynman diagrams
for $\hl\to\gamma\gamma$ are given by the corresponding SM values.
However, there is a new contribution to the one-loop amplitude that arises
from a charged Higgs loop.  But, this contribution is suppressed by 
$\mathcal{O}(v^2/\mhpm^2)$ because
$g\ls{\hl\hp\hm}\sim\mathcal{O}(v)$, and
our assertion is proved.  In addition, the first
non-trivial corrections to decoupling, of $\mathcal{O}(v^2/\mha^2)$,
can easily be computed and arise from two sources.  First, the
contribution of the charged Higgs loop yields a contribution to the 
$\hl\to\gamma\gamma$ amplitude proportional to $g_{\hl H^+H^-}v/\mhpm^2$.
Second, the contribution of the fermion loops are altered due to the
modified $\hl f\bar f$ couplings [see \eq{hlqqlag}], which yield
corrections of $\mathcal{O}(\cbma)\sim\mathcal{O}(v^2/\mha^2)$.
Both corrections enter at the same order.
Note that the contribution of the $W$ loop is also modified, but the
corresponding first order correction is of $\mathcal{O}(\cbma^2)$
[since the $\hl W^+W^-$ coupling is proportional to $\sbma$] and thus
can be neglected.

The above considerations can be generalized to all loop-induced processes
which involve the $\hl$ and SM particles as external states.  As long as
$\lam_i\lsim\mathcal{O}(1)$, the Appelquist-Carazzone decoupling 
theorem~\cite{appelquist}
guarantees that for $\mha\to\infty$, the amplitudes for such processes
approach the corresponding SM values.  The same result also applies
to radiatively-corrected $\hl$ decay rates and cross-sections.

\section{A SM-like Higgs boson without decoupling}\label{sec:threehalf}
 
We have demonstrated above that the decoupling limit (where $\mha^2\gg 
|\lam_i|v^2$) implies that $|\cbma|\ll 1$.  However, the $|\cbma|\ll 1$ 
limit is more general than the decoupling limit.  From
\eq{cbmadec}, one learns that $|\cbma|\ll 1$ implies that either 
(i)~$\mha^2\gg\lambda_A v^2$, and/or (ii) $|\lamhat|\ll 1$
subject to the condition specified by \eq{mhacond}.  Case (i) is the
decoupling limit described in Section~3.  
Although case~(ii) is compatible with $\mha^2\gg\lambda_i v^2$,
which is the true decoupling limit, there is no requirement {\it a priori} 
that $\mha$ be particularly large [as long as \eq{mhacond} is satisfied].  
It is even possible to have $\mha<\mhl$, implying that
all Higgs boson masses are $\lsim\mathcal{O}(v)$, in contrast to the true
decoupling limit.  In this latter case, there does not exist an 
effective low-energy scalar theory consisting of a single Higgs boson.  

Although the tree-level couplings of $\hl$ to vector bosons may appear
to be SM-like, a significant deviation of either the $\hl\overline DD$ or
$\hl\overline UU$ coupling from the corresponding SM value is
possible.  For example, for $|\cbma|\ll 1$, the $\hl$ couplings to
quark pairs normalized to their SM values
[see eqs.~(\ref{cbmadec}), (\ref{qqcouplings1}) and
(\ref{qqcouplings2})] are given by:
\beq \label{largeaqqcouplings}
\hl\overline D D:\quad 1-{\lamhat v^2\tan\beta\over\mha^2-\lam_A v^2}
\,,\qquad\qquad
\hl\overline U U:\quad 1+{\lamhat v^2\cot\beta\over\mha^2-\lam_A v^2}\,.
\eeq
If $\mha\lsim\mathcal{O}(v)$ and $\tanb\gg 1$ [$\cotb\gg 1$], 
then the deviation of the $\hl\overline DD$ [$\hl\overline UU$]
coupling from the corresponding SM value can be significant even
though $|\lamhat|\ll~1$.  A particularly nasty case is one
where the $\hl\overline D D$ [$\hl\overline U U$] coupling is equal in
magnitude but opposite in sign to the corresponding SM 
value~\cite{Ginzburg:2001wj}.\footnote{Note that 
for $|\lamhat|\ll 1$ [{\it i.e.}, for $|\cbma|\ll 1$ with $\mha$ 
arbitrary, where the $\hl VV$ couplings are SM-like], there is
no choice of parameters for which {\it both} the $\hl\overline D D$ and 
$\hl\overline U U$ couplings are equal in magnitude but opposite in sign
relative to the corresponding SM couplings.}
For example, the $\hl \overline DD$ coupling of \eq{largeaqqcouplings}
is equal to $-1$ when
$\tanb\simeq 2[(\mha^2/v^2)-\lam_A]/\lamhat\gg~1$.  Of course, 
the latter corresponds to an isolated point of the parameter space;
it is far more likely that the $\hl\overline D D$
coupling will exhibit a discernible deviation in magnitude from its SM value.
  
Even if the tree-level couplings of $\hl$ to both vector bosons and
fermions appear to be SM-like, radiative corrections
can introduce deviations from SM expectations~\cite{Ginzburg:2001wj}
if $\mha$ is not significantly larger than
$v$.\footnote{Radiative corrections that contribute
to shifts in the coefficients of operators of dimension $\leq 4$ will
simply renormalize the parameters of the scalar potential.
Hence the deviation from the SM of the properties of $\hl$ 
associated with dimension $\leq 4$ operators will 
continue to be suppressed in the limit of the renormalized parameter 
$|\lamhat|\ll 1$.}
For example, consider the amplitude for $\hl\to\gamma\gamma$
(which corresponds to a dimension-five effective operator).
If $\mha\lsim\mathcal{O}(v)$
[implying that $\mhpm\sim\mathcal{O}(v)$] {\it and}
$|\lamhat|\ll 1$ (implying that tree-level couplings of $\hl$ 
approach their SM values), then the charged Higgs boson
loop contribution to the $\hl\to\gamma\gamma$ amplitude will {\it not}
be suppressed.  Hence the resulting amplitude will be shifted from
the SM result, thus revealing that true decoupling has not been achieved,
and the $\hl$ is not the SM Higgs boson~\cite{Ginzburg:2001wj}.

Radiative corrections can also introduce deviations from SM
expectations if the Higgs self-coupling parameters are 
large~\cite{kanemura}.  We can illustrate this in a model in which
$\hl$ is SM-like and all other Higgs bosons are very heavy, and yet
the decoupling limit does not apply.
Consider a model in which $m_{12}^2=\lam_6=\lam_7=0$ and the
Higgs potential parameters are chosen to yield $\mhh=\mha=\mhpm$ and $\cbma=0$.
This can be achieved by taking $m_{11}^2=m_{22}^2$ and\footnote{In this
case, \eqns{minconditionsa}{minconditionsb}
imply that $\tan^2\beta=(\lam_{345}-\lam_1)/(\lam_{345}-\lam_2)$.} 
\beq \label{nondecoupmodel}
\lam_1=\lam_3+{\lam_5 c_{2\beta}\over \cb^2}\,\qquad
\lam_2=\lam_3-{\lam_5 c_{2\beta}\over \sb^2}\,\qquad
\lam_4=\lam_5\,,
\eeq
with $\lam_5<0$ and $-(\lam_1\lam_2)^{1/2}<\lam_{345}<0$
[thereby ensuring that $\mha^2>0$,
$\mhl<\mhh$ and \eq{globalb} are satisfied].  These results are most
easily obtained by using \eqs{inverseA}{inverseD}.  One immediately finds
that $\mhl^2=(\lam_3+\lam_5)v^2$ and $\mhh^2=\mha^2=\mhpm^2=-\lam_5 v^2$.
It is easy to check that $\lamhat=0$ is exact, which yields $\cbma=0$
(since $\lam_{345}<0$ implies that $\mha^2>m_L^2$ and $\mhl^2=m_L^2$
[{\it cf.}~\eqns{cpevenhiggsmasses}{cosbmasq}]), 
and $\lambar=\lamA=\lam_3+\lam_5$.
Note that although $\lamhat=\cbma=0$, \eq{cbmadec} implies that
the ratio
$\lamhat/\cbma=-\lam_{345}=(\mha^2-\mhl^2)/v^2$ can be taken to be 
an arbitrary positive parameter.  This example 
exhibits a model in which the properties of
$\hl$ are indistinguishable from those of 
the SM Higgs boson, but the decoupling limit can never be achieved
(since $m_{12}^2=0$).  One cannot take the masses of the
mass-degenerate $\hh$, $\ha$ and $\hpm$ arbitrarily large 
with $\mhl\sim\mathcal{O}(\mz)$ without taking all the $|\lam_i|$
($i=1,\ldots,5$) arbitrarily large (thereby violating unitarity).
Nevertheless, if one takes the $|\lam_i|$ close to their unitarity
limits, one can find a region of parameter space in which 
$\mhh=\mha=\mhpm\gg\mhl\sim\mathcal{O}(\mz)$.  If only $\hl$ were
observed, it would appear to 
be difficult to distinguish this case from a Higgs sector close to the
decoupling limit.  However, when the
$|\lam_i|$ are large one expects large radiative corrections due to
loops that depend on the Higgs self-couplings.  For example,
the one-loop corrections to the $\hl\hl\hl$ coupling 
(which at tree-level is given by $g_{\hl\hl\hl}=-3\mhl^2/v^2$ when $\cbma=0$)
can deviate by as much as 100\% or more
from the corresponding corrections in the Standard Model
in the above model where $\cbma=0$ and
$\mhh=\mha=\mhpm\gg\mhl\sim\mathcal{O}(\mz)$~\cite{kanemura}.  
More generally, a model with
a light SM-like Higgs boson and all other Higgs bosons
heavy could be distinguished from a Higgs sector near
the decoupling limit only by observing the effects of 
one-loop corrections proportional to the 
(large) Higgs self-coupling parameters.
Such radiative corrections could deviate significantly from
the corresponding loop corrections in the Standard Model.

Two additional examples in which the $|\lamhat|\ll 1$ limit is realized
are given by: 
\ben
\item $\tanb\gg 1$, $\lam_6=\lam_7=0$, and $\mha^2> (\lam_2-\lam_5) v^2$,
and

\item $\lam_1=\lam_2=\lam_{345}$, $\lam_6=\lam_7=0$, and
$\mha^2> (\lam_2-\lam_5) v^2$ \cite{getal}.
\een

The condition on $\mha^2$ in the two cases is required by
\eq{mhacond}.  In case 1, $\lamhat=0$ when $\beta=\pi/2$, whereas in
case 2, $\lamhat=0$ independently of $\tanb$.  In both these cases, it
is straightforward to use \eqns{massmhh}{higgsmasses} to obtain 
\beq
m^2_{\hl,\hh}=\left\{
\begin{array}{l} \lambda_2 v^2 \\[5pt] \mha^2+\lambda_5 v^2  \\
\end{array} \right.
\eeq
Since $m_L^2=\lambda_2 v^2$,
\eq{cosbmasq} yields $\cos(\beta-\alpha)=0$ as expected.

Two special limits of case~2 above 
are treated in \Ref{getal}, where scalar potentials with
$\lam_1=\lam_2=\lam_3=\mp\lam_4=\pm\lam_5>0$ (and
$\lam_6=\lam_7=0$) are considered.  
Assuming that $\mha^2>(\lam_2-\lam_5)v^2$, 
the resulting Higgs spectrum is given by
$\mhpm^2=\mhh^2=\mha^2\pm\mhl^2$ and $\mhl^2=\lambda_1 v^2$ ($\mha^2$ is a
free parameter that depends on $m_{12}^2$).  In the case of $\lam_5>0$,
one has $\mha^2>0$ and it is possible to have a Higgs spectrum 
in which $\ha$ is very light, while the other Higgs bosons
(including~$\hl$) are heavy and approximately degenerate in mass.
In the case of $\lam_5<0$, one has
$\mha^2>2\mhl^2$, and a light $\ha$ would 
imply that all the Higgs bosons of the model are light. 
In both cases $\cbma=0$, and the tree-level
couplings of $\hl$ correspond
precisely to those of the SM Higgs boson [see Section~\ref{sec:three}].
These are clearly very special cases, corresponding to a distinctive form
of the quartic terms of the Higgs potential:
\beq
\calv_4=\half\lam_1\left[(\Phi_1^\dagger
\Phi_1+\Phi_2^\dagger\Phi_2)^2\pm(\Phi_1^\dagger\Phi_2
-\Phi_2^\dagger\Phi_1)^2\right]\,,
\eeq
where the choice of sign corresponds to the sign of $\lam_5$.
Note that $\mathcal{V}_4$ above exhibits a flat direction
if $\lam_5>0$, whereas the scalar potential possesses a globally 
stable minimum if $\lam_5<0$ [see \eq{globalb}].

Next, we examine a region of Higgs parameter space where
$|\sin(\beta-\alpha)|\ll 1$, in which the heavier CP-even Higgs boson,
$\hh$ is SM-like (also considered in \Refs{Ginzburg:2001wj}{getal}).
In this case, the $\hl$ couplings to vector boson pairs 
are highly suppressed.
This is far from the decoupling regime.  Nevertheless,
this region does merit a closer examination, which we now perform.

When $\sbma\to 0$, we have $\beta-\alpha=0$ or $\pi$.  
We shall work to first nontrivial order in a $\sbma$ expansion,
with $\cbma\simeq \pm(1-\half\sbma^2)$.  
Using the results of \eqs{altmha}{altmhh} and \eq{mamthree},
we obtain:\footnote{Note
that \eqns{lamtomsqa}{lamtomsqb} are interchanged under the
transformation $\mhl^2\leftrightarrow\mhh^2$ and $\cbma\leftrightarrow
-\sbma$.  Thus, applying these transformations to
\eqs{decouplimits1}{decouplimits4} yields the results given in
\eqs{otherlimits1}{otherlimits4} with $\cbma=+1$.} 
\beqa
\mha^2 &\simeq &
v^2 \left[\mp{\lamhat\over\sbma}+\lam_A
\pm\nicefrac{3}{2}\lamhat\sbma\right]\,,\label{otherlimits1}\\
\mhl^2 &\simeq &
v^2 \left[\mp{\lamhat\over\sbma}+\lam\pm\half\lamhat\sbma\right]\simeq \mha^2
+(\lambda-\lambda_A\mp\lamhat\sbma)v^2\,,\label{otherlimits2}\\
\mhh^2 &\simeq & v^2(\lambar\pm\lamhat\sbma)\,,\label{otherlimits3}\\
\mhpm^2 &\simeq &
v^2 \left[\mp{\lamhat\over\sbma}+\lam_A+\half\lamF
\pm\nicefrac{3}{2}\lamhat\sbma\right]
=\mha^2+\half\lamF v^2\,.
\label{otherlimits4}
\eeqa
The condition $\mhh>\mhl$ imposes 
the inequality (valid to first order in $\sbma$):
\beq
\mha^2< v^2(\lam_A\pm 2\lamhat\sbma)\,,
\label{mhacond2}
\eeq
[{\it cf.}~\eq{mhaineq}].
Note that \eq{mhacond2} implies that all Higgs squared-masses are 
of $\mathcal{O}(v^2)$.
We may also use \eq{massha} to obtain
\beq
m_{12}^2 \simeq
v^2\sb\cb \left[\mp{\lamhat\over\sbma}+
\lam_{\ha}+\lam_5+\half\lam_6\tb^{-1}+\half\lam_7\tb
\pm\nicefrac{3}{2}\lamhat\sbma\right]\,.
\eeq

We can rewrite \eq{otherlimits1} in another form
[or equivalently use \eqns{exactbma1}{exactbma2} to obtain]:
\beq \label{cbmadec2}
\sin(\beta-\alpha)\simeq  {\mp\lamhat v^2 \over \mha^2-\lambda_A v^2}
\simeq {\pm\lamhat v^2\over \mhh^2-\mhl^2}\,.
\eeq
The $|\sbma|\ll 1$ limit is achieved when $|\lamhat|\ll 1$, subject to
the condition given in \eq{mhacond2}.
Clearly, $\hh$ is SM-like, since if $\sbma\simeq 0$, then the
couplings of $\hh$ to $VV$, $\hh\hh$ and $\hh\hh\hh$ coincide with the
corresponding SM Higgs boson couplings.\footnote{When $\lamhat=0$,
the $\hh$ couplings to $VV$, $\hh\hh$, and $f\bar f$
[see \eq{hhqqlag}] all differ by an overall sign from the corresponding
$\hsm$ couplings if $\cbma=-1$.  
However, this sign is unphysical, since one can eliminate
it with a redefinition $\hl\to -\hl$ and $\hh\to -\hh$, which is equivalent
to replacing $\alpha$ with $\alpha\pm\pi$.}

The couplings of $\hh$ to
fermion pairs are obtained from \eq{yukform} by expanding the Yukawa couplings
of $\hh$ to $\mathcal{O}(\sbma)$
\beqa \label{hhqqlag}
\!\!\!\!\!\!\!\!\!\!\!\!\!\!\!\!\!\!
\call_{\hh QQ} =  &\,- \anti D\, & \left[\pm{1\over v}M_D+\tanb
\left[{1\over v}M_D
-{1\over\sqrt{2}\sb}\left(S_D+iP_D\gamma_5\right)\right]\sbma\right]D
\,\hh\nonumber
\\
 &\,- \anti U\, & \left[\pm{1\over v}M_U-\cotb\left[{1\over v}M_U
-{1\over\sqrt{2}\cb}\left(S_U+iP_U\gamma_5\right)\right]\sbma\right]U
\,\hh\,,
\eeqa
where $\pm$ corresponds to $\cbma=\pm 1$ and $S_D$ and $P_D$ are given by
\eq{sdpd}.  If $|\lamhat|\tanb\ll 1$ or $|\lamhat|\cot\beta\ll 1$, 
then the $\hh f\bar f$ couplings reduce to the corresponding 
$\hsm f\bar f$ couplings.  
However, if $|\lamhat|\ll 1\lsim |\lamhat|\tanb$ [or  
$|\lamhat|\ll 1\lsim |\lamhat|\cot\beta$] when $\tanb\gg 1$ [or 
$\cot\beta\gg 1]$, then the $\hh f\bar f$ couplings can deviate significantly
from the corresponding $\hsm f\bar f $ couplings.  This behavior
is qualitatively different from the decoupling limit, where for fixed
$\lam_i$ and large $\tanb$ [or large $\cot\beta$], one can always choose 
$\mha$ large enough such that the $\hl f\bar f$ couplings approach the
corresponding SM values.  
In contrast, when $|\sbma|\ll 1$, the size of $\mha$ is 
restricted by \eq{mhacond2}, and so there is no guarantee of SM-like 
$\hh f\bar f$ couplings when either $\tanb$ or $\cot\beta$ is large.

Although the tree-level properties of $\hh$ are SM-like when 
$|\lamhat|\ll 1$, deviations can occur for loop-induced 
processes as noted earlier.  Again, the $\hh\to\gamma\gamma$ amplitude
will deviate from the corresponding SM amplitude due to the contribution of
the charged Higgs loop which is not suppressed since 
$\mhpm\sim\mathcal{O}(v)$.  Thus, departures from true decoupling
can in principle be detected for $|\sbma|\ll 1$.

We now briefly examine some model examples in which $|\sbma|\ll 1$ is
realized.  These examples are closely related to the ones previously
considered in the case of $\cbma=0$.  First, consider the model
in which $m_{12}^2=\lam_6=\lam_7=0$ and the
Higgs potential parameters are chosen to yield $\mhl=\mha=\mhpm$ and $\sbma=0$.
This can be achieved by taking $m_{11}^2=m_{22}^2$ and the non-zero 
$\lam_i$ given by \eq{nondecoupmodel} with
$\lam_5<0$ and $\lam_{345}>0$.  In this case,
$\mhh^2=(\lam_3+\lam_5)v^2$ and $\mhl^2=\mha^2=\mhpm^2=-\lam_5 v^2$.
It is easy to check that $\lamhat=0$ is exact and yields $\sbma=0$
(since $\lam_{345}>0$ implies that $\mha^2<m_L^2$ and $\mhh^2=m_L^2$
[{\it cf.}~\eqns{cpevenhiggsmasses}{cosbmasq}]).
Thus, the properties of
$\hh$ are indistinguishable from those of the SM Higgs boson.
However, all the other mass-degenerate Higgs bosons are
lighter than the SM-like Higgs boson, $\hh$.  Thus, one expects that
all Higgs bosons can be observed (once the SM-like Higgs boson is
discovered).  That is, there is little chance of confusing
$\hh$ with the Higgs boson of the Standard Model. 

Two additional examples in which the $|\sbma|\ll 1$ limit is realized
are given by: 
\ben
\item $\tanb\gg 1$, $\lam_6=\lam_7=0$, and $\mha^2< (\lam_2-\lam_5)
v^2$, and

\item $\lam_1=\lam_2=\lam_{345}$, $\lam_6=\lam_7=0$, and
$\mha^2< (\lam_2-\lam_5) v^2$ \cite{getal}.
\een
The condition on $\mha^2$ in the two cases is required by
\eq{mhacond2}.  In case 1, $\lamhat=0$ when $\beta=\pi/2$, whereas in
case 2, $\lamhat=0$ independently of $\tanb$.  In both these cases, it
is straightforward to use \eqns{massmhh}{higgsmasses} to obtain 
\beq
m^2_{\hl,\hh}=\left\{
\begin{array}{l} \mha^2+\lambda_5 v^2  \\[5pt] \lambda_2 v^2 \\
\end{array} \right. \label{hmasses2}
\eeq
Since $m_L^2=\lambda_2 v^2$, it follows from 
\eq{cosbmasq} that $\cbma^2=1$.
Hence, $\sin(\beta-\alpha)=0$, which implies that 
$\hh$ is SM-like.\footnote{Since $\lam_6=\lam_7=0$,
if we additionally set $m_{12}^2=0$, then
we recover the discrete symmetry of the Higgs potential previously
noted in Section~\ref{sec:twohalf}.  Thus, there is no true
decoupling limit in this model.  Moreover, since $\mha^2=-\lambda_5
v^2$ (which implies that $\lambda_5<0$), 
\eq{hmasses2} yields $\mhl=0$, although this result would be 
modified once radiative corrections are included.}

Finally, we note that the SM-like Higgs bosons resulting from
the limiting cases above where $\lamhat=0$ can be easily understood in terms of
the squared-mass matrix entries of \eqns{massmhh}{curlybdef}.
In order to achieve $\cbma=0$ 
or $\sbma=0$, we demand that $\tan 2\beta=\tan 2\alpha$. 
This implies [see \eq{alphadef}]
that the entries in the $\calb^2$ matrix be in the same ratio as the
entries in the term proportional to $\mha^2$ in \eq{massmhh}:
\beq
{2\calm_{12}^2\over \calm_{11}^2-\calm_{22}^2}=\tan 2\beta
\,.\label{eqbma}
\eeq
It is easy to check that 
\beq
\lamhat v^2=\half(\calb_{11}^2-\calb_{22}^2)\sin 2\beta-\calb_{12}^2\cos
2\beta\,.
\eeq
\Eqns{massmhh}{eqbma} immediately imply that $\lamhat=0$ is equivalent
to $\tan 2\beta=\tan 2\alpha$.  Moreover, to determine whether
$\cbma=0$ or $\sbma=0$, simply note that if the sign of 
$\sin 2\alpha/\sin 2\beta$
is negative [positive], then $\cbma=0$ [$\sbma=0$].
In the convention where $\tanb$ is positive, 
it follows that $\sin 2\beta>0$.  Using \eqns{massmhh}{curlybdef},
if the sign of
\beq
\calm_{12}^2=
 \sb\cb\left[(\lamtil-\lam_5)v^2-\mha^2\right]+
v^2(\lam_6\cb^2+\lam_7\sb^2)
\label{signchoice}
\eeq
is negative [positive], then $\cbma=0$ [$\sbma=0$].
One can check that the conditions given by \eqns{mhacond}{mhacond2}
correspond precisely to the negative [positive] sign of $\calm_{12}^2$
[\eq{signchoice}], after imposing $\lamhat=0$.\footnote{It is simplest
to use $\lamhat=0$ to eliminate the quantity
$\lam_1\cb^2-\lam_2\sb^2$ from $\lambda_A$ in 
eqs.~(\ref{mhacond})~and~(\ref{mhacond2}).}
The condition $\lamhat=0$ can be achieved not only 
for appropriate choices of the $\lam_i$ and $\tanb$
in the general 2HDM, but also can be fulfilled
in the MSSM when radiative corrections are 
incorporated (see Section~6).

\section{Decoupling Effects in the MSSM Higgs Sector} \label{sec:five}

The Higgs sector of the MSSM is a CP-conserving
two-Higgs-doublet model, with a Higgs potential whose dimension-four
terms respect supersymmetry and with type-II
Higgs-fermion couplings.
The quartic couplings $\lambda_i$ are given by~\cite{ghsusy}
\beq \label{bndfr}
\lambda_1 =\lambda_2 = -\lamtil=\quarter (g^2+g'^2)\,,\qquad
\lam_4=-\half g^2\,,\qquad \lam_5=\lam_6=\lam_7=0\,.
\eeq
The squared-mass parameters defined in \eq{massdefs} simplify to:
$m_L^2=\mz^2\cos^2 2\beta$, $m_D^2=0$, $m_T^2=\mz^2$ and 
$m_S^2=\mha^2+\mz^2$.  Using \eq{bndfr}, the
invariant coupling parameters defined in
\eqs{lambardef}{lamfdef} and \eqs{lamtdef}{lamvdef} reduce to
\beqa
\lam =-\lam_T=\lam_V &=&\quarter(g^2+g^{\prime 2})\cos^2 2\beta\,,\nonumber \\
\lamhat = -\lam_U &=& \quarter(g^2+g^{\prime 2})\sin 2\beta\cos 2\beta\,,
\nonumber \\
\lam_A &=&\quarter(g^2+g^{\prime 2})\cos 4\beta\,,\nonumber \\
\lamF &=& \half g^2\,.
\eeqa
The results of Section~2 can then be used to obtain the well known
tree-level results:
\beq
\mha^2 = m_{12}^2(\tan\beta+\cot\beta)\,,\qquad\qquad
\mhpm^2 = \mha^2+\mw^2\,,
\eeq
and a neutral CP-even squared-mass matrix given by
\beq 
{\cal M}_0^2 =    \left(
\begin{array}{ll}
\mha^2 \sin^2\beta + m^2_Z \cos^2\beta&
         \qquad  -(\mha^2+m^2_Z)\sin\beta\cos\beta\\
 -(\mha^2+m^2_Z)\sin\beta\cos\beta& \qquad
 \mha^2\cos^2\beta+ m^2_Z \sin^2\beta\end{array}\right)\,,\label{kv}
\eeq
with eigenvalues
\beq
  m^2_{H^0,h^0} = \half \left( \mha^2 + m^2_Z \pm
                  \sqrt{(\mha^2+m^2_Z)^2 - 4m^2_Z \mha^2 \cos^2 2\beta}
                  \; \right)\,,\label{kviii}
\eeq
and the diagonalizing angle $\alpha$ given by
\beq
  \cos 2\alpha = -\cos 2\beta \left( {\mha^2-m^2_Z \over
                  m^2_{H^0}-m^2_{h^0}}\right)\,,\qquad
  \sin 2\alpha = -\sin 2\beta \left( m^2_{H^0} + m^2_{h^0} \over
                   m^2_{H^0}-m^2_{h^0} \right)\,.\label{kix}
\eeq
One can also write
\beq
\cos^2(\beta-\alpha)={\mhl^2(\mz^2-\mhl^2)\over
\mha^2(\mhh^2-\mhl^2)}\,.
\label{cbmasq}
\eeq
In the decoupling limit where $\mha\gg\mz$, the above
formulae yield
\beqa
\mhl^2 &\simeq\ &\mz^2\cos^2 2\beta\,,\qquad\quad
\mhh^2 \simeq\ \mha^2+\mz^2\sin^2 2\beta\,,\nonumber \\[5pt]
\mhpm^2 &=\ & \mha^2+\mw^2\,,\qquad\quad\,\,
\cos^2(\beta-\alpha)\simeq\ {\mz^4\sin^2 4\beta\over 4\mha^4}\,.
\label{largema}
\eeqa
That is, $\mha\simeq\mhh
\simeq\mhpm$, up to corrections of $\mathcal{O}(\mz^2/\mha)$, and
$\cos(\beta-\alpha)=0$ up to corrections of $\mathcal{O}(\mz^2/\mha^2)$.

It is straightforward to work out all the tree-level Higgs couplings,
both in general and in the decoupling limit.  Since the Higgs-fermion
couplings follow the type-II pattern, the Higgs-fermion Yukawa
couplings are given by \eq{yukform} with $\eta_1^U=\eta_2^D=0$.  
However, one-loop radiative corrections can lead in some cases to
significant shifts from the tree-level couplings.  It is of interest to
examine how the approach to the decoupling limit is affected
by the inclusion of radiative corrections.

First, we note that in some cases, one-loop effects mediated by loops
of supersymmetric particles can generate a deviation from Standard
Model expectations, even if $\mha\gg\mz$ where the corrections to
the decoupling limit are negligible.  As a simple example, if squarks
are relatively light, then squark loop contributions to the $\hl\to
gg$ and $\hl\to\gamma\gamma$ amplitudes can be 
significant~\cite{Djouadi:1996pb}.  Of
course, in the limit of large squark masses, the contributions of the
supersymmetric loops
decouple as well~\cite{Dobado:1997up}.  
Thus, in the MSSM, there are two separate
decoupling limits that must be analyzed.  For simplicity, we assume
henceforth that supersymmetric particle masses are large (say of order
1~TeV), so that supersymmetric loop effects of the type just mentioned
are negligible.

The leading contributions to the radiatively-corrected Higgs
couplings arise in two ways.  First, the radiative corrections
to the CP-even Higgs squared-mass matrix results in a shift 
of the CP-even Higgs mixing angle $\alpha$ from its tree-level value. 
That is, the dominant Higgs propagator corrections 
can to a good approximation be absorbed into an effective 
(``radiatively-corrected'') mixing angle $\alpha$~\cite{hffsusyprop}.
In this approximation, we can write:
\beq \label{calmatrix} 
{\cal M}^2\equiv
\left( \matrix{{\cal M}_{11}^2 &  {\cal M}_{12}^2 \crr {\cal
M}_{12}^2 &  {\cal M}_{22}^2 } \right) ={\cal M}_0^2+\delta {\cal
M}^2\,, 
\eeq 
where the tree-level contribution ${\cal M}_0^2$ was
given in \eq{kv} and $\delta {\cal M}^2$ is the contribution from
the radiative corrections.
Then, $\cos(\beta-\alpha)$ is given by
\beqa \label{eq:cosbma}
\cos(\beta-\alpha)&=&{(\mathcal{M}_{11}^2-\mathcal{M}_{22}^2)\sin 2\beta
-2\mathcal{M}_{12}^2\cos 2\beta\over
2(m_H^2-m_h^2)\sin(\beta-\alpha)} \nonumber \\[5pt]
&=&{m_Z^2\sin 4\beta+
({\delta\mathcal{M}}_{11}^2-{\delta\mathcal{M}}_{22}^2)\sin 2\beta
-2{\delta\mathcal{M}}_{12}^2\cos 2\beta\over
2(m_H^2-m_h^2)\sin(\beta-\alpha)}\,.
\eeqa
Using tree-level Higgs couplings with $\alpha$ replaced by its 
renormalized value provides a useful first approximation to 
the radiatively-corrected Higgs couplings.

Second, contributions from the
one-loop vertex corrections to tree-level Higgs-fermion couplings can
modify these couplings in a significant way, especially
in the limit of large $\tan\beta$.  In particular, although the
tree-level Higgs-fermion coupling follow the type-II pattern, when
radiative corrections are included, all possible dimension-four
Higgs-fermion couplings are generated.
These results can be summarized by an effective
Lagrangian that describes the coupling of
the neutral Higgs bosons to the third generation
quarks:
\begin{equation} \label{susyyuklag}
        -\mathcal{L}_{\rm eff} = \left[
        (h_b + \delta h_b) \bar b_R b_L \Phi_1^{0\ast} 
        + (h_t + \delta h_t) \bar t_R t_L \Phi_2^0 \right]
        + \Delta h_t \bar t_R t_L \Phi_1^0 
        + \Delta h_b \bar b_R b_L \Phi_2^{0 \ast}
        + {\rm h.c.}\,,
\end{equation}
resulting in a modification of the tree-level relation between
$h_t$ [$h_b$] and $m_t$ [$m_b$] as
follows~\cite{deltamb,deltamb1,hffsusyqcd,deltamb2}:
\beqa
        m_b &=& \frac{h_b v}{\sqrt{2}} \cos\beta
        \left(1 + \frac{\delta h_b}{h_b}
        + \frac{\Delta h_b \tan\beta}{h_b} \right)
        \equiv\frac{h_b v}{\sqrt{2}} \cos\beta
        (1 + \Delta_b)\,, \label{byukmassrel} \\[5pt]
        m_t &=& \frac{h_t v}{\sqrt{2}} \sin\beta
        \left(1 + \frac{\delta h_t}{h_t} + \frac{\Delta
        h_t\cot\beta}{h_t} \right)
        \equiv\frac{h_t v}{\sqrt{2}} \sin\beta
        (1 + \Delta_t)\,. \label{tyukmassrel}
\eeqa
The dominant contributions to $\Delta_b$ are $\tan\beta$-enhanced,
with $\Delta_b\simeq (\Delta h_b/h_b)\tan\beta$; for
$\tan\beta\gg 1$, $\delta h_b/h_b$ provides a small correction to
$\Delta_b$.   [In the same limit, $\Delta_t\simeq\delta h_t/h_t$, with
the additional contribution of $(\Delta h_t/h_t)\cot\beta$ providing a
small correction.]

From \eq{susyyuklag} we can obtain the couplings of the physical 
neutral Higgs
bosons to third generation quarks.  The resulting interaction
Lagrangian is of the form:
\beq \label{linthff}
{\cal L}_{\rm int} =  -\sum_{q=t,b} \left[g_{\hl q\bar q}\hl q \bar{q} +
 g_{\hh q\bar q}\hh q \bar{q}-
 i g_{\ha q\bar q} A \bar{q} \gamma_5 q\right]\,.
\eeq
Using \eqns{byukmassrel}{tyukmassrel}, one obtains~\cite{chlm,howmar}:
\beqa
 g_{\hl b\bar b} &= & -{m_b\over v}{\sin\alpha \over \cos\beta}
\left[1+{1\over 1+\Delta_b}\left({\delta h_b\over h_b}-
\Delta_b\right)\left( 1 +\cot\alpha \cot\beta \right)\right]\,,
\label{hlbb} \\[5pt]
 g_{\hh b\bar b} &= & {m_b\over v}{\cos\alpha \over \cos\beta}
\left[1+{1\over 1+\Delta_b}\left({\delta h_b\over h_b}-
\Delta_b\right)\left( 1 -\tan\alpha \cot\beta \right)\right]\,,
\label{hhbb} \\[5pt]
 g_{\ha b\bar b} &= & {m_b\over v}\tan\beta
\left[1+{1\over (1+\Delta_b)\sin^2\beta}\left({\delta h_b\over h_b}-
\Delta_b\right)\right]\,,
\label{habb} \\[5pt]
 g_{\hl t\bar t} & = & {m_t\over v}{\cos\alpha \over \sin\beta}
\left[1-{1\over 1+\Delta_t}{\Delta h_t\over h_t}
(\cot\beta+ \tan\alpha)\right]\,,
\label{hltt} \\[5pt]
 g_{\hh t\bar t} &= & {m_t\over v}{\sin\alpha \over \sin\beta}
\left[1-{1\over 1+\Delta_t}{\Delta h_t\over h_t}
(\cot\beta-\cot\alpha)\right]\,,
\label{hhtt} \\[5pt]
 g_{\ha t\bar t} &= & {m_t\over v}\cot\beta
\left[1-{1\over 1+\Delta_t}{\Delta h_t\over h_t}(\cot\beta+
\tan\beta)\right]\,.
\label{hatt}
\eeqa

We now turn to the decoupling limit.  First
consider the implications for the radiatively-corrected value of
$\cos(\beta-\alpha)$.  
Since $\delta\mathcal{M}^2_{ij}\sim {\mathcal O}(m_Z^2)$,
and $m_H^2-m_h^2=m_A^2+\mathcal{O}(m_Z^2)$, one finds~\cite{chlm}
\begin{equation} \label{cosbmadecoupling}
        \cos(\beta-\alpha)=c\left[{m_Z^2\sin 4\beta\over
        2m_A^2}+\mathcal{O}\left(m_Z^4\over m_A^4\right)\right]\,,
\end{equation}
in the limit of $\mha\gg\mz$, where
\begin{equation} \label{cdef}
        c\equiv 1+{{\delta\mathcal{M}}_{11}^2-{\delta\mathcal{M}}_{22}^2\over
        2m_Z^2\cos 2\beta}-{{\delta\mathcal{M}}_{12}^2\over m_Z^2\sin
        2\beta}\,.
\end{equation}
The effect of the radiative corrections has been to modify the
tree-level definition of $\lamhat$:
\beq \label{lamhatrad}
\lamhat v^2= c\mz^2\sin 2\beta\cos 2\beta\,.
\eeq
\Eq{cosbmadecoupling} exhibits the expected decoupling behavior
for $m_A\gg m_Z$.
However, \eqns{eq:cosbma}{cosbmadecoupling} exhibit another way in which
$\cos(\beta-\alpha)=0$ can be achieved---simply choose the
MSSM parameters (that govern the Higgs mass radiative
corrections) such that the numerator of \eq{eq:cosbma} vanishes.
That is,
\begin{equation}
        2 m_Z^2 \sin 2\beta =
        2\, \delta \mathcal{M}^2_{12}
        - \tan 2\beta
        \left(\delta \mathcal{M}^2_{11} - \delta \mathcal{M}^2_{22} \right)\,.
        \label{eq:tanbetadecoup}
\end{equation}
This condition is equivalent to $c=0$, and thus 
corresponds precisely to the case of $\lamhat=0$ discussed at the
beginning of Section~5.  Although $\lamhat\neq 0$ at tree-level, the
above analysis shows that $|\lamhat|\ll 1$ can arise due to the
effects of one-loop radiative corrections that approximately cancel the
tree-level result.\footnote{The one-loop corrections arise from the
exchange of supersymmetric particles, whose contributions can be
enhanced for certain MSSM parameter choices.  
One can show that the 
two-loop corrections are subdominant, so that the approximation scheme
is under control.}  In particular,
\eq{eq:tanbetadecoup} is independent of the value of $m_A$.
Typically,
\eq{eq:tanbetadecoup} yields a solution at large $\tan\beta$.  That
is, by
approximating $\tan 2\beta\simeq -\sin 2\beta \simeq -2/ \tan \beta$,
one can determine
the value of $\beta$ at which $\lamhat\simeq 0$~\cite{chlm}:
\begin{equation} \label{earlydecoupling}
\tan \beta\simeq \frac{2m_Z^2-
\delta\mathcal{M}_{11}^2+\delta \mathcal{M}_{22}^2}
{ \delta\mathcal{M}_{12}^2}\,.
\end{equation}
Hence, there exists a value of $\tanb$ (which depends on the choice
of MSSM parameters) where
$\cos(\beta-\alpha)\simeq 0$ independently of
the value of $m_A$.  If $\mha$ is not much larger than
$\mz$, then $\hl$ is
a SM-like Higgs boson outside the decoupling regime.\footnote{For
large $\tanb$ and $\mha\lsim\mathcal{O}(\mz)$, one finds that
$\sinbma\simeq 0$, implying that
$\hh$ is the SM-like Higgs boson, as discussed
in Section~5.}  
Of course,
as explained in Section~5, this SM-like Higgs boson can be
distinguished in principle
from the SM Higgs boson by measuring its decay rate  
to two photons and looking for a deviation from SM predictions.

Finally, we analyze the radiatively-corrected Higgs-fermion couplings
[\eqs{hlbb}{hatt}] in the decoupling limit.  Here it is useful to note
that for $\mha\gg\mz$,
\begin{equation}  \label{cotalf}
\cot\alpha = -\tan\beta - \frac{2\mz^2}{\mha^2} \tan\beta \cos
2\beta + {\cal O}\left(\frac{\mz^4}{\mha^4}\right)\,.
\end{equation}
Applying this result to \eqns{hlbb}{hltt}, it follows 
that in the decoupling limit, $g_{\hl q\bar q}=
g_{\hsm q\bar q}=m_q/v$.
Away from the decoupling limit,
the Higgs couplings to down-type fermions can deviate significantly
from their tree-level values due to enhanced radiative corrections at
large $\tan\beta$ [where $\Delta_b\simeq\mathcal{O}(1)$].
In particular, because $\Delta_b\propto\tan\beta$, the leading
one-loop radiative correction to $g_{\hl b\bar b}$ is of
$\mathcal{O}(\mz^2\tanb/\mha^2)$, which formally decouples only when
$\mha^2\gg\mz^2\tanb$.  This behavior is called {\it delayed
decoupling} in \Ref{loganetal}, although this phenomenon can
also occur in a more general 2HDM (with tree-level couplings), as
noted previously in Section~4 [below \eq{sdpd}].

\section{Discussion and Conclusions}

In this paper, we have studied the decoupling limit of a general
CP-conserving
two-Higgs-doublet model.  In this limit, the lightest Higgs boson of
the model is a CP-even neutral Higgs scalar ($\hl$) 
with couplings identical
to those of the SM Higgs boson.  Near the decoupling
limit, the first order corrections for the Higgs couplings to gauge
and Higgs bosons, the Higgs-fermion Yukawa couplings and the Higgs
cubic and quartic self-couplings have also been obtained.  
These results exhibit a
definite pattern for the deviations of the $\hl$ couplings from those
of the SM Higgs boson.  In particular, the rate of 
the approach to decoupling depends on the particular Higgs coupling
as follows:
\beqa
   {g^2_{hVV}\over g^2_{\hsm VV}} & \simeq &
1-{\lamhat^2 v^4\over m_A^4}\,, \label{sqcoup1} \\[5pt]      
{g^2_{hhh}\over g^2_{\hsm\hsm\hsm}} &\simeq & 1-{6\lamhat^2 v^2\over
    \lam\mha^2}\,, \\[5pt]
   {g^2_{htt}\over g^2_{\hsm tt}} & \simeq & 1+{2\lamhat v^2\cot\beta
\over\mha^2}\left(1-\xi_t\right)\,,  \\[5pt]
  {g^2_{hbb}\over g^2_{\hsm bb}} & \simeq & 1-{2\lamhat v^2\tan\beta
\over \mha^2}\left(1-\xi_b\right)\,, \label{sqcoup4}
\eeqa    
where $\xi_t$ and $\xi_b$ reflect the terms proportional to $S$ and
$P$ in \eq{hlqqlag}.  Thus, the
approach to decoupling is fastest for the $\hl$ couplings to
vector bosons and slowest for the couplings to down-type 
(or up-type) quarks if $\tanb>1$ (or $\tanb<1$).
We may apply the above results to the MSSM (see Section 6).
Including the leading $\tanb$-enhanced
radiative corrections, 
$\xi_b=v\Delta h_b/(\sqrt{2}\sb m_b)=
\Delta_b/(\sb^2(1+\Delta_b))$ [whereas $\xi_t\ll 1$ can be neglected]
and $\lamhat$ is given by \eq{lamhatrad}.
Plugging into \eqs{sqcoup1}{sqcoup4}, one reproduces the results
obtained in \Ref{chlm}.

Although the results of this paper were derived from a tree-level
analysis of couplings, these results can also be applied to the
radiatively-corrected couplings that multiply operators of
dimension-four or less. An example of this was given in
Section 6, where we showed how the decoupling limit applies to the
radiatively-corrected Higgs-fermion Yukawa couplings.   
In particular, near the decoupling limit one can neglect radiative
corrections that are generated by the exchange of heavy Higgs bosons.
These contributions are suppressed by a loop factor in addition to the
suppression factor of $\mathcal{O}(v^2/\mha^2)$ and thus are smaller
than corrections to tree-level Higgs couplings
that enter at first order in $\cbma$.  This should be contrasted
with loop-induced Higgs couplings ({\it e.g.}, $\hl\to\gamma\gamma$
which is generated by a dimension-five effective operator), where the
corrections of $\mathcal{O}(\cbma)$
to tree-level Higgs couplings that appear in the one-loop
amplitude and the effects of a heavy
Higgs boson loop are both of $\mathcal{O}(v^2/\mha^2)$ [in
addition to the overall one-loop factor].
Consequently, both contributions are equally important in
determining the overall correction to the
loop-induced Higgs couplings due to the
departure from the decoupling limit.

If a neutral Higgs boson, $\hl$, is discovered at a future collider, it may
turn out that its couplings are close to those expected of the SM
Higgs boson.  The challenge for future experiments is then to
determine whether the observed state is the SM Higgs boson, or whether
it is the lowest lying scalar state of a non-minimal Higgs 
sector~\cite{burgess}.
If the latter, then it is likely that the additional scalar states of
the model are heavy, and the decoupling limit applies.
In this case, it is possible that the heavier scalars cannot be
detected at the LHC or at an $e^+e^-$ linear collider (LC) with a
center-of-mass energy in the range of 350---800~GeV.  
Moreover, it may not be
possible to distinguish between the $\hl$ and the SM-Higgs boson at
the LHC.  However, the measurements of Higgs observables at the LC
can provide sufficient precision to observe deviations from SM Higgs
properties at the few percent level.  In this case, one can begin to
probe deep into the decoupling regime~\cite{boudjema}.

In this paper, we also clarified a Higgs parameter regime in which $\hl$
possesses SM-like couplings to vector bosons but where
$\mha^2\lsim\mathcal{O}(v^2)$ and the decoupling
limit does not apply (see Section~5).  
In this case, the couplings of $\hl$ to fermion
pairs can deviate significantly from the corresponding SM
Higgs-fermion couplings if either $\tanb$ or $\cotb$ is large.
Moreover, the masses of $\hh$, $\ha$ and $\hpm$ are not particularly
large, and all scalars would be accessible at the LHC and/or the LC.

The discovery of the Higgs boson will be a remarkable achievement.
Nevertheless, the lesson of the decoupling limit is that a 
SM-like Higgs boson provides very little information about the nature of
the underlying electroweak symmetry-breaking dynamics.
It is essential to find evidence
for departures from SM Higgs predictions.  
Such departures can
reveal crucial information about the existence of a
non-minimal Higgs sector.
Precision Higgs measurements can also provide critical tests
of possible new physics beyond the Standard Model.  As an example,
in the MSSM, deviations in Higgs couplings from the decoupling limit 
can yield indirect information about the MSSM parameters.  
In particular, at large $\tanb$ the sensitivity to MSSM parameters 
may be increased due to enhanced radiative corrections.
The decoupling limit is both a curse and an opportunity.  If nature
chooses the Higgs sector parameters to lie deep in the decoupling
regime, then it may not be possible to distinguish the observed $\hl$
from the SM Higgs boson.  On the other hand, given sufficient
precision of the measurements of $\hl$ branching ratios and
cross-sections~\cite{howmar}, it may be possible to observe a 
small but statistically significant
deviation from SM expectations, and provide a first glimpse 
of the physics responsible for electroweak symmetry breaking.

\acknowledgments

We are grateful to Scott Thomas and Chung Kao 
for their contributions at the initial stages of this project.
We also appreciate a number of useful conversations with
Maria Krawczyk.  Finally, we 
would like to thank Marcela Carena, Heather Logan 
and Steve Mrenna for various insights concerning the supersymmetric
applications based on the results of this paper.
This work was supported in part by the U.S. Department of Energy.


\appendix 
\section{An alternative parameterization of the 2HDM scalar potential}
\label{app:A}

In this Appendix, we give the translation of the parameters
of eq.~(\ref{pot}) employed in this paper to the parameters
employed in the Higgs Hunter's Guide (HHG) \cite{hhg}. While the HHG
parameterization was useful for some purposes
({\it e.g.}, the scalar potential minimum is explicitly exhibited), it 
obscures the decoupling limit.

In the HHG parameterization, the most general 2HDM scalar potential,
subject to a discrete symmetry $\Phi_1\to -\Phi_1$ that is only
softly violated by dimension-two terms, is given by\footnote{In the 
HHG, the $V_i$ and $\Lambda_i$ are denoted by
$v_i$ and $\lambda_i$, respectively.  In \eq{hhgpot}, we employ the
former notation in order to distinguish between the HHG
parameterization and the notation of \eqns{pot}{potmin}.}
\beqa \calv&=&
\Lam_1\left(\Phi_1^\dagger\Phi_1-{V_1^2}\right)^2
+\Lam_2\left(\Phi_2^\dagger\Phi_2-{V_2^2}\right)^2
+\Lam_3\left[\left(\Phi_1^\dagger\Phi_1-{V_1^2}\right)
  +\left(\Phi_2^\dagger\Phi_2-{V_2^2}\right)\right]^2 \nonumber\\
&+&\Lam_4\left[(\Phi_1^\dagger\Phi_1)(\Phi_2^\dagger\Phi_2)-
  (\Phi_1^\dagger\Phi_2)(\Phi_2^\dagger\Phi_1)\right]
+\Lam_5\left[\Re(\Phi_1^\dagger\Phi_2)-{V_1V_2}\cos\xi\right]^2
\nonumber\\
&+&\Lam_6\left[\Im(\Phi_1^\dagger\Phi_2)-{V_1V_2}\sin\xi\right]^2
\nonumber\\
&+&\Lam_7\left[\Re(\Phi_1^\dagger\Phi_2)-{V_1V_2}\cos\xi\right]
\left[\Im(\Phi_1^\dagger\Phi_2)-{V_1V_2}\sin\xi\right]\,,
\label{hhgpot}
\eeqa
where the $\Lambda_i$ are real parameters.\footnote{In \eq{hhgpot}
we include the $\Lam_7$ term that was left out in the hardcover
edition of the HHG.  See the erratum 
that has been included in the paperback edition of the HHG
(Perseus Publishing, Cambridge, MA, 2000).}
The $V_{1,2}$ are related
to the $v_{1,2}$ of eq.~(\ref{potmin}) by $V_{1,2}=v_{1,2}/\sqrt 2$.
The conversion from these $\Lam_i$ to
the $\lam_i$ and $m_{ij}^2$ of eq.~(\ref{pot}) is:
\beqa
&&\lam_1=2(\Lam_1+\Lam_3)\,,\nonumber\\
&&\lam_2=2(\Lam_2+\Lam_3)\,,\nonumber\\
&&\lam_3=2\Lam_3+\Lam_4\,,\nonumber\\
&&\lam_4=-\Lam_4+\half(\Lam_5+\Lam_6)\,,\nonumber\\
&&\lam_5=\half(\Lam_5-\Lam_6-i\Lam_7)\,,\nonumber\\
&&\lam_6=\lam_7=0\nonumber\\
&&m_{11}^2=-2V_1^2\Lam_1-2(V_1^2+V_2^2)\Lam_3\,,\nonumber\\
&& m_{22}^2=-2V_2^2\Lam_2-2(V_1^2+V_2^2)\Lam_3\,,\nonumber\\
&& m_{12}^2=V_1V_2(\Lam_5\cos\xi-i\Lam_6\sin\xi-\ihalf e^{i\xi}\Lam_7)\,.
\label{conversion}
\eeqa
Excluding $\lambda_6$ and $\lambda_7$, the scalar potential
[\eqns{pot}{hhgpot}] are fixed by ten real parameters.  The
CP-conserving limit of \eq{hhgpot} is most easily obtained by setting
$\xi=0$ and $\Lam_7=0$.  In the CP-conserving limit, it is easy to
invert \eq{conversion} and solve for the $\Lam_i$ ($i=1,\ldots, 6$).
The result is:
\beqa \label{convertback}
&&\Lam_1=\half\left[\lam_1-\lamtil+2m_{12}^2/(v^2\sb\cb)\right]\,,\nonumber\\
&&\Lam_2=\half\left[\lam_2-\lamtil+2m_{12}^2/(v^2\sb\cb)\right]\,,\nonumber\\
&&\Lam_3=\half\left[\lamtil-2m_{12}^2/(v^2\sb\cb)\right]\,,\nonumber\\
&&\Lam_4=2m_{12}^2/(v^2\sb\cb)-\lam_4-\lam_5\,,\nonumber\\[5pt]
&&\Lam_5=2m_{12}^2/(v^2\sb\cb)\,,\nonumber\\[5pt]
&&\Lam_6=2m_{12}^2/(v^2\sb\cb)-2\lam_5\,,
\eeqa
where $\lamtil\equiv\lam_3+\lam_4+\lam_5$ and $v^2\sb\cb=2V_1 V_2$.

\section{Conditions for CP conservation in the two-Higgs doublet
  model}
\label{app:B}

\def\phid{\Phi_1^\dagger}
\def\phii{\Phi_2}
\def\phi{\Phi_1}
\def\phiid{\Phi_2^\dagger}
\def\To{\Rightarrow}
First, we derive the conditions such that the Higgs sector does
not exhibit explicit CP violation.\footnote{For another approach,
in which invariants are employed to identify basis-independent
conditions for CP violation in the Higgs sector, see
\Refs{lavoura}{branco}.}   
It is convenient to adopt a convention
in which one of the vacuum expectation values, say $v_1$ is real and
positive.\footnote{Due to the U(1)-hypercharge symmetry of the theory,
it is always possible to make a phase rotation on the scalar fields such
that $v_1>0$.}  This still leaves one additional phase redefinition for
the Higgs doublet fields.  If there is no explicit CP violation,
it should be possible to choose the phases of the Higgs fields
so that there are no explicit phases in the Higgs potential parameters
of \eq{pot}. If we consider
$\phid\phii\to e^{-i\eta}\phid\phii$, then the $\eta$-dependent 
terms in $\calv$ are given by
\beqa
&&
\calv\ni 
-m_{12}^2e^{-i\eta}\phid\phii+\half\lam_5e^{-2i\eta}\left(\phid\phii\right)^2
\nonumber\\
&&
+\lam_6e^{-i\eta}\left(\phid\phi\right)\left(\phid\phii\right)
+\lam_7e^{-i\eta}\left(\phiid\phii\right)\left(\phid\phii\right) +
{\rm h.c.}
\label{etapot}
\eeqa
Let us write
\beq
m_{12}^2=|m_{12}^2|e^{i\theta_m}\,,\qquad
\lam_{5,6,7}=|\lam_{5,6,7}|e^{i\theta_{5,6,7}} \,.
\eeq
Then, all explicit parameter phases are removed if 
\beq
\theta_m-\eta=n_m\pi\,,\quad \theta_5-2\eta=n_5\pi\,,\quad
\theta_{6,7}-\eta=n_{6,7}\pi\,,
\label{thetaform}
\eeq
where $n_{m,5,6,7}$ are integers.
Writing $\eta=\theta_m-n_m\pi$ from the first condition of \eq{thetaform}, and
substituting into the other conditions, gives
\beqa
\theta_5-2\theta_m&=& (n_5-2n_m)\pi\quad\To {\Im}[(m_{12}^2)^2\lam_5^*]=0\,,
\label{cpi}\\
\theta_6-\theta_m&=& (n_6-n_m)\pi\quad\To {\Im}[m_{12}^2\lam_6^*]=0\,,
\label{cpii}\\
\theta_7-\theta_m&=&(n_7-n_m)\pi\quad\To {\Im}[m_{12}^2\lam_7^*]=0\,.
\label{cpiii}
\eeqa
\Eqs{cpi}{cpiii} constitute the conditions for the absence of explicit
CP violation in the (tree-level) Higgs sector.
A useful convention is one in which $m_{12}^2$ is real (by a suitable
choice of the phase $\eta$).  It then follows that $\lambda_5$,
$\lambda_6$ and $\lambda_7$ are also real.
Henceforth, we shall assume that all parameters in the scalar potential
are real.

Let us consider now the conditions for the absence of spontaneous CP
violation.\footnote{Similar considerations can be found in
refs.~\cite{branco,tdlee,vsb} and \cite{Dubinin:2002nx}.}
Let us write $\vev{\phid\phii}=\half v_1v_2e^{i\xi}$ with $v_1$ and
$v_2$ real and positive and $0\leq\xi\leq\pi$.
The $\xi$-dependent terms in $\calv$ are given by
\beq
\calv\ni
-m_{12}^2v_1v_2\cos\xi+\quarter\lam_5v_1^2v_2^2\cos2\xi+
\half\lam_6 v_1^3v_2\cos\xi
+\half\lam_7 v_2^3v_1\cos\xi\,,
\eeq
which yields
\beq
{\partial \calv\over \partial
  \cos\xi}=-m_{12}^2v_1v_2+\lam_5v_1^2v_2^2\cos\xi
+\half\lam_6 v_1^3v_2 + \half\lam_7  v_2^3v_1
\eeq
and
\beq
{\partial^2 \calv\over \partial(\cos\xi)^2}=\lam_5v_1^2v_2^2\,.
\eeq
Spontaneous CP violation occurs when $\xi\neq 0, \pi/2$ or $\pi$
at the potential minimum.  That is,   
$\lam_5>0$ and there exists a CP-violating solution to
\beq
\cos\xi={m_{12}^2-\half \lam_6 v_1^2-\half \lam_7 v_2^2\over \lam_5v_1v_2}\,.
\eeq
Thus, we conclude that the criterion for spontaneous CP violation (in
a convention where all parameters of the scalar potential are real) is
\beq
0\neq \left|m_{12}^2-\half\lam_6 v_1^2-\half\lam_7v_2^2\right|<
    \lam_5 v_1v_2\qquad {\rm and}~\lambda_5>0\,.
\label{scpcondition}
\eeq
Otherwise, the minimum of the potential occurs either at $\xi=0$,
$\pi/2$ or
$\pi$ and CP is conserved.\footnote{The CP-conserving minimum 
corresponding to $\xi=0$ or $\xi=\pi$ does
not in general correspond to an extremum in $V(\cos\xi)$.
Specifically, for $\lambda_5<0$, the extremum corresponds to a
maximum in $\calv$, while for $\lambda_5>0$, the extremum corresponding to a
minimum of $\calv(\cos\xi)$ arises for $|\cos\xi|>1$.  In both cases,
when restricted to
the physical region corresponding to $|\cos\xi|\leq 1$, the minimum
of $\calv(\cos\xi)$ is attained on the boundary, $|\cos\xi|=1$.}
The case of $\xi=\pi/2$ is singular and 
arises when $m_{12}^2=\half\lam_6 v_1^2+\half\lam_7
v_2^2$ and $\lam_5>0$.\footnote{Note that the 
case of $\xi=\pi/2$ arises automatically in the case of the discrete 
symmetry discussed in Section~\ref{sec:twohalf},
$m_{12}^2=\lam_6=\lam_7=0$,
when $\lam_5>0$.}  
It is convenient to choose
a convention where $\vev{\Phi_1^0}$ is real and
$\vev{\Phi_2^0}$ is pure imaginary.  One must then
re-evaluate the Higgs mass eigenstates. 
As shown in \Ref{chinese},
the neutral Goldstone boson is now a linear combination of 
${\rm Im}~\Phi_1^0$ and ${\rm Re}~\Phi_2^0$,
while the physical CP-odd scalar,
$\ha$ corresponds to the orthogonal combination.  The two CP-even
Higgs scalars are orthogonal linear combinations of 
${\rm Re}~\Phi_1^0$ and ${\rm Im}~\Phi_2^0$.  Most of the results of
this paper do not apply for this case without substantial revision.
Nevertheless, it is clear that the decoupling limit 
($\mha^2\gg\lam_i v^2$) does not exist due to the condition on
$m_{12}^2$.  

We shall not consider the $\xi=\pi/2$ model further in
this paper.  Then, if
the parameters of the scalar potential are real and if
there is no spontaneous CP-violation, then it is always possible to choose 
the phase $\eta$ in \eq{etapot} so that the potential minimum
corresponds to $\xi=0$.\footnote{In particular, if $\xi=\pi$, simply
choose $\eta=\pi$, which corresponds to changing the overall sign of 
$\Phi_1^\dagger\Phi_2$.  This is equivalent to redefining the parameters
$m_{12}^2\to -m_{12}^2$, $\lambda_6\to -\lambda_6$ and $\lambda_7\to
-\lambda_7$.}  In this convention,
\beqa
m_{12}^2-\half\lam_6 v_1^2 -\half\lam_7 v_2^2 &\geq& \lambda_5 v_1 v_2
\qquad {\rm for}~\lambda_5> 0\label{nocpconditioni} \,,\\[5pt]
m_{12}^2-\half\lam_6 v_1^2 -\half\lam_7 v_2^2 &\geq& 0 \qquad
\;\qquad {\rm for}~\lambda_5\leq 0\,,
\label{nocpconditionii}
\eeqa
where \eq{nocpconditioni} follows from \eq{scpcondition}, and
\eq{nocpconditionii} is a consequence of
the requirement that $\calv(\xi=0)\leq \calv(\xi=\pi)$.
Since $\xi=0$ and both $v_1$ and $v_2$ are real and positive, this 
convention corresponds to the one chosen below \eq{v246}.
Note that if we rewrite \eq{massha} as~\footnote{
Under the assumption that $v_1$ and $v_2$ are positive, \eq{massha}
implicitly employs the convention in which $\xi=0$.}
\beq
\mha^2 ={v^2\over v_1 v_2}\left[m_{12}^2-\lambda_5 v_1 v_2
-\half\lambda_6 v_1^2-\half\lambda_7 v_2^2\right]\,,\label{masshacp}
\eeq
it follows that if $\lambda_5> 0$, then the condition $\mha^2\geq
0$ is equivalent to \eq{nocpconditioni}.
However, if  $\lambda_5\leq 0$,  then \eq{nocpconditionii} implies that
$\mha^2\geq |\lambda_5| v^2$.

\section{A singular limit: $\mhl=\mhh$}
\label{app:X}

By definition, $\mhl\leq\mhh$.  The limiting case of $\mhl=\mhh$ is
special and requires careful treatment in some cases.   
For example, despite
the appearance of $\mhh^2-\mhl^2$ in the denominator of \eq{cosbmasq},
one can show that $0\leq\cbma^2\leq 1$.  To prove this, we first write
\beq \label{cbmaeq2}
\cbma^2={1\over 2}\left[1-{m_S^2-2m_L^2\over
\sqrt{m\ls{S}^4-4\mha^2\mlsq-4m\ls{D}^4}}\right]\,.
\eeq
Next, we use \eq{massdefs} to explicitly compute:
\beq \label{essfour}
m\ls{S}^4-4\mha^2\mlsq-4m\ls{D}^4=\mha^4-2\mha^2\left[(\calb^2_{22}
-\calb^2_{11})\ctwob+2\calb^2_{12}\stwob\right]+
(\calb^2_{11}-\calb^2_{22})^2 +4[\calb^2_{12}]^2\,,
\eeq
and
\beq \label{esstwo}
(m_S^2-2m_L^2)^2=m\ls{S}^4-4\mha^2\mlsq-4m\ls{D}^4
-\left[(\calb^2_{11}-\calb^2_{22})\stwob-2\calb^2_{12}\ctwob\right]^2\,.
\eeq
Note that \eq{essfour}, viewed as a quadratic function of
$\mha^2$ (of the form $Am_A^4+Bm_A^2+C$), is non-negative if
$B^2-4AC=\left[(\calb^2_{11}-\calb^2_{22})\stwob-2\calb^2_{12}
\ctwob\right]^2\geq 0$.  It then follows from \eq{cbmaeq2}
that $0\leq\cbma^2\leq 1$ if
\beq \label{inequality}
(m_S^2-2m_L^2)^2\leq m\ls{S}^4-4\mha^2\mlsq-4m\ls{D}^4\,,
\eeq
a result which is manifestly true [see \eq{esstwo}].

We now turn to the case of $\mhl=\mhh$.  This can arise if and only if
the CP-even Higgs squared-mass matrix (in any basis) 
is proportional to the unit
matrix.  From \eq{massmhh}, it then
follows that:
\beq \label{equalcond}
\calb^2_{11}-\calb^2_{22}=\mha^2\ctwob\,,\qquad\quad 
2\calb^2_{12}=\mha^2\stwob\,.
\eeq
where $\mhl^2=\mhh^2=\calb^2_{11}+\mha^2\sb^2=\calb^2_{22}+\mha^2\cb^2$.
Alternatively, from \eq{cpevenhiggsmasses}, the condition for $\mhl=\mhh$
is given by
$m\ls{S}^4-4\mha^2\mlsq-4m\ls{D}^4\equiv A\mha^4+B\mha^2+C=0$.
However, one must check that this quadratic equation possesses a 
positive (real) solution for $\mha^2$.  Noting the discussion above
\eq{inequality}, such a solution can exist if and only
if $B^2-4AC=0$, which is indeed consistent with \eq{equalcond}.
Of course, the results of \eq{equalcond} are
not compatible with the decoupling limit,
since it is not possible to have $\mhl=\mhh$ and $\mha^2\gg |\lam_i|v^2$.

If we take $B^2-4AC=0$ but keep $\mha$ arbitrary,
then \eq{cbmaeq2} yields
\begin{equation}
\cbma^2=\left\{
\begin{array}{rcc}
           0\,, & \quad\text{if} & m_L^2<\half m_S^2\,, \\
           1\,, & \quad\text{if} & m_L^2>\half m_S^2\,.
        \end{array} \right.
\end{equation}
For $m_L^2=\half m_S^2$, we have $\mhl^2=\mhh^2=\half m_S^2$, and the
angle $\alpha$ is not well-defined.  In this case, one cannot
distinguish between $\hl$ and $\hh$ in either production or decays, and the
corresponding squared-amplitudes should be (incoherently) added in all
processes.  It is easy to check that the undetermined angle $\alpha$
that appears in the relevant Higgs couplings would then drop out in any
such sum of squared-amplitudes.  The singular point of parameter space
corresponding to $\mhl=\mhh$ will not be considered further in this paper.

\section{Relations among Higgs potential parameters and masses}
\label{app:C}

It is useful to express the physical Higgs masses in terms of the
parameters of the scalar potential [\eq{pot}]. 
First, inserting \eqns{massmhh}{curlybdef} into
\eq{diagn} and examining the diagonal elements
yields the CP-even Higgs boson squared-masses:
\beqa
\mhl^2&=&\mha^2\cbma^2
+v^2\bigl[\lam_1\cb^2\sa^2+\lam_2\sb^2\ca^2
-2\lamtil\ca\cb\sa\sb+\lam_5\cbma^2\cr
&&\phantom{\mha^2\cbma^2+v^2\bigl[\lam_1}-2\lam_6\cb\sa\cbpa+
2\lam_7\sb\ca\cbpa\bigr]\,,
\label{mhlform}
\\
\mhh^2&=&\mha^2\sbma^2
+v^2\bigl[\lam_1\cb^2\ca^2+\lam_2\sb^2\sa^2
+2\lamtil\ca\cb\sa\sb+\lam_5\sbma^2\cr
&&\phantom{\mha^2\cbma^2+v^2\bigl[\lam_1}+2\lam_6\cb\ca\sbpa+
2\lam_7\sb\sa\sbpa\bigr]\,,
\label{mhhform}
\eeqa
while the requirement that the off-diagonal entries
in \eq{diagn} are zero yields
\beqa
\mha^2\,\sbma\cbma&=&\half v^2\bigl[\stwoa(-\lam_1\cb^2+\lam_2\sb^2)
+\lamtil\stwob\ctwoa-2\lam_5 \sbma\cbma\cr
&&\phantom{{v^2\over 2}\bigl[\stwoa}
+2\lam_6\cb c_{\beta+2\alpha}
+2\lam_7\sb s_{\beta+2\alpha}\bigr]\,,
\label{mhaform}
\eeqa
where $\lamtil\equiv\lam_3+\lam_4+\lam_5$.
We can now eliminate $\mha^2$ from \eqns{mhlform}{mhhform}
and \eqns{massha}{mamthree} using the result of \eq{mhaform}.
This yields equations for the other three physical Higgs boson
squared-masses and the scalar potential mass parameter $m_{12}^2$
in terms of the Higgs scalar quartic couplings
\beqa
{\mhl^2\over v^2}\sbma &=&-\lam_1\cb^3\sa+\lam_2\sb^3\ca+\half
\lamtil\cbpa\stwob \nonumber \\
&&+\lam_6\cb^2(\cb\ca-3\sb\sa)+\lam_7\sb^2(3\cb\ca-\sb\sa)\,,
\label{lamtomsqa}\\[5pt]
{\mhh^2\over v^2}\cbma &=&\lam_1\cb^3\ca+\lam_2\sb^3\sa+\half
\lamtil\sbpa\stwob \nonumber \\
&&+\lam_6\cb^2(3\sb\ca+\cb\sa)+\lam_7\sb^2(\sb\ca+3\cb\sa)\,,
\label{lamtomsqb}\\[5pt]
{2\mhpm^2\over v^2}\sbma\cbma 
&=&-\stwoa(\lam_1\cb^2-\lam_2\sb^2)+\lamtil\stwob\ctwoa
-(\lam_4+\lam_5)\sbma\cbma\nonumber\\
&&+2\lam_6\cb c_{\beta+2\alpha}+2\lam_7\sb s_{\beta+2\alpha} \,,
\label{lamtomsqc}\\[5pt]
{2m_{12}^2\over v^2}\sbma\cbma &=&-\half\stwob
\stwoa(\lam_1\cb^2-\lam_2\sb^2)+\half\lamtil\stwob^2\ctwoa\nonumber\\
&&+\lam_6\cb^2\left[3\cb\sb\ctwoa-\ca\sa(1+2\sb^2)\right]
+\lam_7\sb^2\left[3\sb\cb\ctwoa+\ca\sa(1+2\cb^2)\right]\,.\nonumber \\[6pt]
&&\phantom{equation}
\label{lamtomsqd}
\eeqa
Note that \eq{lamtomsqc} is easily derived by inserting \eq{mhaform}
into \eq{mamthree}. A related useful result is
easily derived from \eqns{mhaform}{lamtomsqb}:
\beqa
{(\mha^2-\mhh^2)\over v^2}\sbma&=&\half\stwob\left(
-\lam_1\ca\cb+\lam_2\sa\sb
+\lamtil\cbpa\right)-\lam_5\sbma\nonumber\\
&&+\lam_6\cb\left[\cb\cbpa-2\sb^2\ca\right]
+\lam_7\sb\left[\sb\cbpa+2\cb^2\sa\right]\,.
\label{hahhdif}
\eeqa
It is remarkable that the left hand side of 
\eq{hahhdif} is proportional only to
$\sbma$ ({\it i.e.}, the factor of $\cbma$ has canceled).  As a result, in
the decoupling limit where $\cbma\to 0$, we see that
$\mha^2-\mhh^2={\mathcal{O}}(v^2)$.

The expressions given in \eqs{mhaform}{lamtomsqc} are quite
complicated.  These results simplify considerably when expressed in
terms of $\lam$, $\lamhat$ and $\lam_A$ [\eqs{lambardef}{lamadef}]:
\beqa 
\mha^2&=&v^2\left[\lam_A+\lamhat\left({\sbma\over\cbma}-
{\cbma\over\sbma}\right)\right]\,, \label{altmha} \\[5pt]
\mhl^2&=&v^2\left[\lam-{\lamhat\,\cbma\over\sbma}\right]\,, \label{altmhl}\\[5pt]
\mhh^2&=&v^2\left[\lam+{\lamhat\,\sbma\over\cbma}\right]\,. \label{altmhh}
\eeqa
One can then rewrite
\eq{hahhdif} as
\beq \label{mhhmhadiff}
\mhh^2-\mha^2=v^2\left[\lam-\lam_A+{\lamhat\cbma\over\sbma}\right]\,.
\eeq

We can invert \eqs{mhaform}{lamtomsqd} and solve for any five of the scalar
potential parameters in terms of the physical Higgs masses and the
remaining three 
undetermined variables~\cite{boudjema,Dubinin:1998nt,Santos:2001tp}.
It is convenient to solve for
$\lambda_1,\ldots,\lambda_5$ in terms of $\lambda_6$, $\lambda_7$,
$m_{12}^2$ and the Higgs masses.  
We obtain:
\beqa
\lam_1&=& {\mhh^2\ca^2+\mhl^2\sa^2-m_{12}^2\tb\over v^2\cb^2}
-\thalf\lam_6\tb+\half\lam_7\tb^3\,,
\label{inverse1}\\
\lam_2&=& {\mhh^2\sa^2+\mhl^2\ca^2-m_{12}^2\tb^{-1}\over v^2\sb^2}
+\half\lam_6\tb^{-3}-\thalf\lam_7\tb^{-1}\,,
\label{inverse2}\\
\lam_3&=& { (\mhh^2-\mhl^2){\ca\sa}
+2\mhpm^2\sb\cb-{m_{12}^2 }
\over v^2\sb\cb}-\half\lam_6\tb^{-1}-\half\lam_7\tb\,
\label{inverse3}\\
\lam_4&=& {(\mha^2-2\mhpm^2)\sb\cb+m_{12}^2  \over v^2\sb\cb}
-\half\lam_6\tb^{-1}-\half\lam_7\tb\,,
\label{inverse4}\\
\lam_5&=&{m_{12}^2-\mha^2\sb\cb \over v^2\sb\cb}
-\half\lam_6\tb^{-1}-\half\lam_7\tb\,.
\label{inverse5}
\eeqa
In addition, the minimization
conditions of \eqns{minconditionsa}{minconditionsb} reduce to:
\beqa
m_{11}^2&=&
-{1\over 2\cb}\left(\mhh^2\ca\cbma-\mhl^2\sa\sbma\right)+m_{12}^2\tb\,,
\label{minconds1}
\\
m_{22}^2&=&
-{1\over 2\sb}\left(\mhl^2\ca\sbma+\mhh^2\sa\cbma\right)+m_{12}^2\tb^{-1}\,.
\label{minconds2}
\eeqa
Note that 
$\lam_6$ and $\lam_7$ do not appear when $m_{11}^2$ and $m_{22}^2$ 
are expressed
entirely in terms of $m_{12}^2$ and physical Higgs masses.

In some cases, it proves more convenient to eliminate $m_{12}^2$ in
favor of $\lambda_5$ using \eq{inverse5}.  
The end result is:
\beqa
\lam_1&=& {\mhh^2\ca^2+\mhl^2\sa^2-\mha^2\sb^2\over
  v^2\cb^2}-\lam_5\tb^2-2\lam_6\tb \,,\label{inverseA}\\
\lam_2&=& {\mhh^2\sa^2+\mhl^2\ca^2-\mha^2\cb^2\over
  v^2\sb^2}-\lam_5\tb^{-2}-2\lam_7\tb^{-1}\,,\label{inverseB}\\
\lam_3&=& {(\mhh^2-\mhl^2)\sa\ca+(2\mhpm^2-\mha^2)\sb\cb\over
v^2\sb\cb}-\lam_5-\lam_6\tb^{-1}-\lam_7\tb \,,\label{inverseC}\\
\lam_4&=&{2(\mha^2-\mhpm^2)\over v^2}+\lam_5\,,\label{inverseD}
\eeqa
and
\beqa
m_{11}^2&=&
-{1\over 2\cb}\left(\mhh^2\ca\cbma-\mhl^2\sa\sbma\right)+(\mha^2+\lam_5
v^2) \sb^2+\half v^2(\lam_6\sb\cb+\lam_7\sb^2\tb)\,,\nonumber \\
\label{minconds3}
\\
m_{22}^2&=&
-{1\over
  2\sb}\left(\mhl^2\ca\sbma+\mhh^2\sa\cbma\right)+(\mha^2+\lam_5v^2)\cb^2
+\half v^2(\lam_6\cb^2\tb^{-1}+\lam_7\sb\cb)\,. \nonumber \\
\label{minconds4}
\eeqa

Using \eqs{altmha}{altmhh}, one may obtain simple expressions for
$\lam$, $\lamhat$ and $\lam_A$ [\eqs{lambardef}{lamadef}] in terms of the
neutral Higgs squared-masses:
\beqa
\lam\, v^2&=&\mhl^2\,\sbmaii+\mhh^2\,\cbmaii\,,\label{lammasses}\\[5pt]
\lamhat\, v^2 &=& (\mhh^2-\mhl^2)\,\sbma\,\cbma\,,\label{lamhatmasses}\\[5pt]
\lamA\,
v^2&=&\mha^2+(\mhh^2-\mhl^2)\,(\cbmaii-\sbmaii)\,,\label{lamamasses}\\[5pt]
\lamF\, v^2&=& 2(\mhpm^2-\mha^2)\,,\label{lamfmasses}
\eeqa
where we have also included an expression for $\lamF\equiv\lam_5-\lam_4$
in terms of the Higgs squared-masses [see \eq{mamthree}].
Thus, four of the the invariant coupling parameters can be
expressed in terms of
the physical Higgs masses and the basis-independent quantity $\beta-\alpha$
(see Appendix~\ref{app:C2}).

Finally, we note that \eqns{lamhatmasses}{lamamasses} 
also yield a simple expression for $\beta-\alpha$, which plays such a
central role in the decoupling limit.  We find two forms that are
noteworthy:
\beq \label{exactbma1}
\tan\,[2(\beta-\alpha)]={-2\lamhat v^2\over \mha^2-\lamA v^2}\,,
\eeq
and
\beq \label{exactbma2}
\sin\,[2(\beta-\alpha)]={2\lamhat v^2\over\mhh^2-\mhl^2}\,.
\eeq
Indeed, if $\lamhat=0$ then either $\cbma=0$ or $\sbma=0$ as discussed
in Section~5.  For $\lamhat\neq 0$, the condition
$\mhh>\mhl$ implies that $\lamhat\sbma\cbma>0$.  This
inequality, when applied to \eq{altmha}, imposes the following
constraint on $\mha$
\beq \label{mhaineq}
v^2\left[\lamA-{2\lamhat\,\cbma\over\sbma}\right]<\mha^2 <
v^2\left[\lamA+{2\lamhat\,\sbma\over\cbma}\right]\,.
\eeq
In addition, we require that $\mha^2\geq 0$.

The expressions for the Higgs masses [\eqs{altmha}{altmhh}] and
$\beta-\alpha$ [\eq{exactbma1} or (\ref{exactbma2})] are especially
useful when considering the approach to the decoupling limit, where
$|\cbma|\ll 1$.  For example, \eqs{altmha}{altmhh} reduce in this
limit to the results of \eqs{decouplimits1}{decouplimits3}.
Moreover,
$\sin [2(\beta-\alpha)]\simeq -\tan [2(\beta-\alpha)]\simeq 2\cbma$, and
\eqns{exactbma1}{exactbma2} reduce to the results given by \eq{cbmadec}.
The corresponding results in limiting case of $|\sbma|\ll 1$ 
treated in Section~5 are also similarly obtained.

\section{Invariant combinations of the Higgs scalar potential parameters}
\label{app:C2}

In the most general 2HDM model, there is no distinction between the
two $Y=1$ complex doublets, $\Phi_1$ and $\Phi_2$.  In principle, one
could choose any two orthogonal linear combinations of $\Phi_1$
and $\Phi_2$ ({\it i.e.}, choose a new basis for the scalar doublets),
and construct the scalar sector Lagrangian with respect to the new
basis.  Clearly, the parameters of \eq{pot}, $m_{ij}^2$ and the
$\lambda_i$, would all be modified, along with $\alpha$ and $\beta$.
However, there exists seven invariant combinations of the $\lambda_i$
that are independent of basis choice~\cite{inprep}.  These are: $\lambar$,
$\lamhat$, $\lamA$, $\lamF$ 
defined in \eqs{lambardef}{lamfdef}, and $\lamT$,
$\lamU$ and $\lamV$ defined in \eqs{lamtdef}{lamvdef}.
In addition, the combination
$\beta-\alpha$ is clearly basis independent.  Thus, all physical Higgs
masses and Higgs self-couplings can be expressed in terms of the above
invariant coupling parameters and $\beta-\alpha$.  In
Appendix~\ref{app:C}, we have already shown how to express the Higgs
masses in terms of the invariant parameters.  
In Appendices~\ref{app:D} and \ref{app:E}
we also exhibit the three-Higgs and four-Higgs couplings in terms of the
invariant parameters.\footnote{The Higgs couplings to vector bosons
depend only on $\beta-\alpha$ [see \eqs{vvcoup}{littletable}].
The Higgs couplings to
fermions in the Type-III model (in which both up and down-type
fermions couple to both Higgs doublets) can also be written in terms
of invariant parameters.  However, one would then have to identify the
appropriate invariant combinations of the Higgs-fermion Yukawa coupling 
parameters~\cite{inprep}, $\eiui$ and $\eidi$ [see \eq{etayuks}].}

To obtain expressions for the Higgs self-couplings in terms of
invariant parameters, one must invert the relations between the
$\lam_i$ and the invariant coupling parameters.  The end result is:
\beqa \label{invparms}
\!\!\!\!\!\!\!\!\!\!\!\lam_1&=&\cb^2(1+3\sb^2)\lam+2\stwob(\cb^2\,\lamhat
       +\sb^2\,\lamU)-\half s_{2\beta}^2(2\lamA-\lamT)
       +\sb^4\lamV\,,\nonumber\\[5pt]
\!\!\!\!\!\!\!\!\!\!\!\lam_2 &=&\sb^2(1+3\cb^2)\lam-2\stwob(\sb^2\,\lamhat
       +\cb^2\,\lamU)-\half s_{2\beta}^2(2\lamA-\lamT)+\cb^4\lamV\,,
\nonumber\\[5pt]
\!\!\!\!\!\!\!\!\!\!\!\lamtil &=&(2c_{2\beta}^2-\cb^2\sb^2)\lam
      -3\stwob\ctwob(\lamhat-\lamU)-(c_{2\beta}^2-2\cb^2\sb^2)(2\lamA-\lamT)
      +\nicefrac{3}{4}s_{2\beta}^2\lamV\,,\nonumber\\[5pt]
\!\!\!\!\!\!\!\!\!\!\!\lam_5 &=& (c_{2\beta}^2+\cb^2\sb^2)\lam
      -\stwob\ctwob(\lamhat-\lamU)-c_{2\beta}^2\lamA
      +\nicefrac{1}{4}s_{2\beta}^2(\lamV-2\lamT)\,,\nonumber\\[5pt]
\!\!\!\!\!\!\!\!\!\!\!\lam_6 &=& \half\stwob(3\sb^2-1)\lam
      -\cb\cthreeb\lamhat-\sb\sthreeb\lamU+\half\stwob\ctwob(2\lamA-\lamT)
      -\half\sb^2\stwob\lamV\,,\nonumber \\[5pt]
\!\!\!\!\!\!\!\!\!\!\!\lam_7 &=& \half\stwob(3\cb^2-1)\lam
       -\sb\sthreeb\lamhat-\cb\cthreeb\lamU-\half\stwob\ctwob(2\lamA-\lamT)
       -\half\cb^2\stwob\lamV\,,
\eeqa
and $\lam_4=\lam_5-\lamF$.

The significance of the invariant coupling parameters is most evident
in the so-called Higgs basis of \Ref{branco}, in which only the neutral
component of one of the two Higgs doublets (say, the first one)
possesses a vacuum expectation value.  Let us denote the two Higgs
doublets in this basis by $\Phi_a$ and $\Phi_b$.  Then, after a rotation
from the $\Phi_1$--$\Phi_2$ basis by an angle $\beta$,
\beqa
\Phi_a&=&\phm\Phi_1\cosb+\Phi_2\sinb\,,\nonumber \\
\Phi_b&=&-\Phi_1\sinb+\Phi_2\cosb\,,
\eeqa
one obtains
\beq \label{abbasis}
\Phi_a=\left(\begin{array}{c}
G^+ \\ {1\over\sqrt{2}}\left(v+\varphi_a^0+iG^0\right)\end{array}
\right)\,,\qquad
\Phi_b=\left(\begin{array}{c}
H^+ \\ {1\over\sqrt{2}}\left(\varphi_b^0+i\ha\right)\end{array}
\right)\,,
\eeq
where $\varphi_a^0$ and $\varphi_b^0$ are related in the CP-conserving
model to the CP-even neutral Higgs bosons by:
\beqa
\hh&=&\varphi_a^0\cosbma-\varphi_b^0\sinbma\,, \label{hbasis}\\
\hl&=&\varphi_a^0\sinbma+\varphi_b^0\cosbma\,. \label{Hbasis}
\eeqa
Here, we see that $\beta-\alpha$ is the invariant angle that characterizes
the direction of the CP-even mass eigenstates (in the
two-dimensional Higgs ``flavor'' space)
relative to that of the vacuum expectation value.

In the Higgs basis, the
corresponding values of $\lam_1,\cdots,\lam_7$ are easily evaluated by
putting $\beta=0$ in \eq{invparms}.  Thus, the scalar potential takes
the following form:
\beqa
\mathcal{V}&=& m_{aa}^2\Phi_a^\dagger\Phi_a+m_{bb}^2\Phi_b^\dagger\Phi_b
-[m_{ab}^2\Phi_a^\dagger\Phi_b+{\rm h.c.}]\nonumber\\[8pt]
&&\quad\!\!\!\! +\half\lambda(\Phi_a^\dagger\Phi_a)^2
+\half\lamV(\Phi_b^\dagger\Phi_b)^2
+(\lamT+\lamF)(\Phi_a^\dagger\Phi_a)(\Phi_b^\dagger\Phi_b)
+(\lambda-\lamA-\lamF)(\Phi_a^\dagger\Phi_b)(\Phi_b^\dagger\Phi_a)
\nonumber\\[8pt]
&&\quad\!\!\!\! +\left\{\half(\lambda-\lamA)(\Phi_a^\dagger\Phi_b)^2
-\big[\lamhat\,(\Phi_a^\dagger\Phi_a)
+\lamU(\Phi_b^\dagger\Phi_b)\big]
\Phi_a^\dagger\Phi_b+{\rm h.c.}\right\}\,, \label{pothbasis}
\eeqa
where three new invariant quantities are revealed:
\beqa
m_{aa}^2&=&m_{11}^2\cb^2+m_{22}^2\sb^2-[m_{12}^2+(m_{12}^*)^2]\sb\cb\,,\\
m_{bb}^2&=&m_{11}^2\sb^2+m_{22}^2\cb^2+[m_{12}^2+(m_{12}^*)^2]\sb\cb\,,\\
m_{ab}^2&=&(m_{11}^2-m_{22}^2)\sb\cb+m_{12}^2\cb^2-(m_{12}^*)^2\sb^2\,.
\eeqa
In the CP-conserving theory where $m_{12}^2$ is real, 
the corresponding potential minimum 
conditions [\eqs{minconditionsa}{minconditionsb}] simplify to:
\beq \label{potcond}
m_{aa}^2=-\half v^2\lambda\,,\qquad m_{ab}^2=-\half v^2\lamhat\,,
\eeq
with no constraint on $m_{bb}^2$.  In fact, $m_{bb}^2$ is related to
$\mha^2$:
\beqa
\mha^2&=& {\rm Tr}~m^2+\half v^2(\lambda+\lamT)\nonumber \\[4pt]
&=& m_{bb}^2+\half v^2\lamT\,,
\eeqa
after imposing the potential minimum condition [\eq{potcond}].  It is
convenient to trade the free parameter $m_{bb}^2$ for $\beta-\alpha$.
Using the results of \eqns{exactbma1}{exactbma2}, it follows that
\beq
\tan[2(\beta-\alpha)]={2\lamhat\over \lamA-\half\lamT-m_{bb}^2/v^2}\,,
\eeq
where the sign of $\sin[2(\beta-\alpha)]$ is equal to the sign of
$\lamhat$.

It is now straightforward to obtain the three-Higgs and four-Higgs
couplings in terms of the invariant coupling parameters and
$\beta-\alpha$, by inserting \eqs{abbasis}{Hbasis} into \eq{pothbasis}.

\section{Three-Higgs vertices in the two-Higgs doublet model}
      \label{app:D}

In this Appendix, we list the Feynman rules for the three-point Higgs
interaction in the most general CP-conserving
two-Higgs doublet extension of the Standard Model.
The Feynman rule for the $ABC$ vertex is denoted by 
$ig\ls{ABC}$.\footnote{To obtain $g\ls{ABC}$, multiply the coefficient
of $ABC$ that appears in the interaction Lagrangian by the appropriate
symmetry factor $n!$, where $n$ is the number of identical particles
at the vertex.  Note that $H^+$ and $H^-$ are not considered identical.}
For completeness, $R$-gauge Feynman rules
involving the Goldstone bosons ($\gpm$ and $\go$)
are also listed.

The Feynman rules are obtained from the scalar potential by
multiplying the corresponding coefficients of $\mathcal{V}$ by $-i$
times the appropriate symmetry factor.
To obtain the three-Higgs couplings in terms of $\beta-\alpha$ and
the invariant coupling parameters, we
insert \eqs{abbasis}{Hbasis} into \eq{pothbasis}, and
identify the terms that are cubic in the Higgs boson fields.
The resulting three-point Higgs couplings (which are proportional to
$v\equiv 2\mw/g$) are given by:
\beqa \label{invdefghaa}
g\ls{\hl\ha\ha} &=&
   {-v}\bigl[\lamT\sbma-\lamU\cbma\bigr]\,,\nonumber \\[4pt]
g\ls{\hh\ha\ha} &=&  
   {-v}\bigl[\lamT\cbma+\lamU\sbma\bigr]\,,\nonumber \\[4pt]
g\ls{\hl\hh\hh} &=& {3v}\bigl[
   \lam\sbma\left(-\nicefrac{2}{3}+\cbmaii\right)+\lamhat\cbma(1-3\sbmaii)
    \nonumber \\
&&\qquad\qquad    +(2\lamA-\lamT)\sbma\left(\nicefrac{1}{3}-\cbmaii\right)  
    +\lamU\sbmaii\cbma\bigr]\,,\nonumber \\[4pt]
g\ls{\hh\hl\hl} &=& {3v}\bigl[
   \lam\cbma\left(-\nicefrac{2}{3}+\sbmaii\right)-\lamhat\sbma(1-3\cbmaii)
    \nonumber \\
&&\qquad\qquad   +(2\lamA-\lamT)\cbma\left(\nicefrac{1}{3}-\sbmaii\right)  
    -\lamU\cbmaii\sbma\bigr]\,,\nonumber \\[4pt]
g\ls{\hl\hl\hl} &=& {-3v}\bigl[
    \lam\sbma(1+\cbmaii)-3\lamhat\cbma\sbmaii-(2\lamA-\lamT)\sbma\cbmaii
    -\lamU c^3_{\beta-\alpha}\bigr]\,,\nonumber \\[4pt]
g\ls{\hh\hh\hh} &=& {-3v}\bigl[
    \lam\cbma(1+\sbmaii)+3\lamhat\sbma\cbmaii-(2\lamA-\lamT)\cbma\sbmaii
    +\lamU s^3_{\beta-\alpha}\bigr]\,,\nonumber \\[4pt]
g\ls{\hl\hp\hm} &=&
   {-v}\bigl[(\lamT+\lamF)\sbma-\lamU\cbma\bigr]\,,\nonumber \\[4pt]
g\ls{\hh\hp\hm} &=&
   {-v}\bigl[(\lamT+\lamF)\cbma+\lamU\sbma\bigr]\,.
\eeqa
\begin{table}[t!]
\begin{center}
\caption{Three-Higgs vertex Feynman rules in the approach to the decoupling
limit are given by $ig_{ABC}=iv(X_{ABC}+Y_{ABC}\,\cbma)$, where the
coefficients $X$ and $Y$ are listed below. \label{hhhdecouptab}}
\vspace{.2in}
\begin{tabular}{|l|c|c|} \hline
$ABC$ & $X_{ABC}$ & $Y_{ABC}$ \rule{0in}{3ex} \\[1ex] \hline
$\hl\hl\hl$ & $-3\lam$ &  \rule{0in}{3ex}
$9\lamhat$ \rule{0in}{3ex} \\[4pt]
$\hl\hl\hh$ & $-3\lamhat$ & $\lam+2(\lamT-2\lam_A)$ \\[4pt]
$\hl\hh\hh$ & $2(\lam_A-\lam)-\lamT$ &  $3(\lamU-2\lamhat)$ \\[4pt]
$\hl\ha\ha$ & $-\lamT$  &  $\lamU$  \\[4pt]
$\hl\hp\hm$ & $-\lamT-\lamF$ & $\lamU$ \\[4pt]
$\hh\hh\hh$ & $-3\lamU$ & $6(\lam_A-\lam)-3\lamT$ \\[4pt]
$\hh\ha\ha$ & $-\lamU$ &   $-\lamT$ \\[4pt]
$\hh\hp\hm$ & $-\lamU$ & $-\lamT-\lamF$ \\[1ex] \hline
\end{tabular}
\end{center}
\end{table}

In the approach to the
decoupling limit, the three-Higgs vertices simplify
considerably as exhibited in Table~\ref{hhhdecouptab}.  Here, we have
listed all the cubic couplings in the form:
\beq
g_{ABC}= v(X_{ABC}+Y_{ABC}\,\cbma)\,,
\eeq
where the coefficients $X$ and $Y$ are given in terms of various
linear combinations of the invariant coupling parameters.
These results follow trivially from \eq{invdefghaa}.

The couplings involving the Goldstone bosons are given by
\beqa
g\ls{\hl\go\go} &=&g\ls{\hl\gp\gm}=
   v\bigl[\lamhat\cbma-\lam\sbma\bigr]\,,\nonumber \\[3pt]
g\ls{\hh\go\go} &=& g\ls{\hh\gp\gm}=
    -v\bigl[\lamhat\sbma+\lam\cbma\bigr]\,,\nonumber \\[3pt]
g\ls{\hl\ha\go} &=& v\bigl[\lamhat\sbma-(\lam-\lamA)\cbma\bigr]\,,
\nonumber \\[3pt]
g\ls{\hh\ha\go} &=&v\bigl[\lamhat\cbma+(\lam-\lamA)\sbma\bigr] 
\,,\nonumber \\[3pt]
g\ls{\hl\hpm \gmp} &=&
v\bigl[\lamhat\sbma-(\lam-\lamA-\half\lamF)\cbma\bigr]\,,\nonumber \\[3pt]
g\ls{\hh\hpm \gmp} &=&
v\bigl[\lamhat\cbma+(\lam-\lamA-\half\lamF)\sbma\bigr]\,,\nonumber \\[3pt]
g\ls{\ha\hpm \gmp} &=& \pm\nicefrac{i}{2} v\lamF\,.
\eeqa
In the rule for the $\ha \hpm \gmp$ vertex,
the sign corresponds to $\hpm$ entering the vertex and $\gpm$ leaving
the vertex.

One can also express the three-Higgs vertices in terms of the Higgs
masses by using \eqs{lammasses}{lamfmasses}.  
The Feynman rules for the three-point Higgs vertices that involve Goldstone
bosons then take on rather simple forms:
\beqa
g\ls{\hl\go\go} &=&g\ls{\hl\gp\gm}=
   {-g\over 2\mw} \mhl^2\sbma\,,\nonumber \\[3pt]
g\ls{\hh\go\go} &=& g\ls{\hh\gp\gm}=
   {-g\over 2\mw} \mhh^2\cbma\,,\nonumber \\[3pt]
g\ls{\hl\ha\go} &=& {-g\over 2\mw}(\mhl^2-\mha^2)\cbma\,,
\nonumber \\[3pt]
g\ls{\hh\ha\go} &=& {g\over 2\mw}(\mhh^2-\mha^2)\sbma
\,,\nonumber \\[3pt]
g\ls{\hl\hpm \gmp} &=&
{g\over 2\mw}(\mhpm^2-\mhl^2)\cbma\,,\nonumber \\[3pt]
g\ls{\hh\hpm \gmp} &=&
{-g\over 2\mw}(\mhpm^2-\mhh^2)\sbma\,,\nonumber \\[3pt]
g\ls{\ha\hpm \gmp} &=&
{\pm ig\over 2\mw}(\mhpm^2-\mha^2)\,.
\label{goldstonerules}
\eeqa
The cubic couplings of the physical Higgs bosons, expressed in terms of the
Higgs masses, are more complicated.  For example, let us first
compute $g_{\hl\hl\hl}$ in terms of $\lam_1,\cdots,\lam_7$:
\beq
g\ls{\hl\hl\hl} = {3v}\bigl[
   \lambda_1\sa^3\cb-\lambda_2\ca^3\sb+\lamtil
   \sa\ca\cab 
-\lambda_6\sa^2\big(3\ca\cb-\sa\sb\big)
+\lambda_7\ca^2\big(3\sa\sb-\ca\cb\big)\bigr]\,.
\eeq
This can then be re-expressed in terms of the Higgs masses using
\eqs{inverseA}{inverseD}.  The end result is~\cite{boudjema}
\beq \label{hhhalt}
g_{\hl\hl\hl}=-3v\left[{\mhl^2\sbma\over
    v^2}+\left({\mhl^2-\mha^2-\lam_5 v^2\over v^2\sb\cb}\right)
    \cbma^2\cbpa +\left(\lam_6{\sa\over\sb}-\lam_7{\ca\over\cb}\right)
    \cbma^2\right]\,.
\eeq
Note that the
decoupling limit result [\eq{hhhdecoup}] follows easily
after using \eq{decouplimits1} to obtain the $\mathcal{O}(\cbma)$
correction.  We have also exhibited $g_{\hl\hp\hm}$ in \eq{gcubic}.
Expressions for the other three-Higgs couplings in
terms of the Higgs masses can be found in \Ref{boudjema} (see also
\Ref{Dubinin:1998nt} for the case of $\lam_6=\lam_7=0$
and \Ref{santosetal} for other special cases).  However, in the most
general case, such expressions are less useful.  Finally, 
using \eq{lamfmasses} we note the relations
\beqa
v[g\ls{\hl\hp\hm}-g\ls{\hl\ha\ha}]&=&
   -2(\mhpm^2-\mha^2)\sba\,,\nonumber \\[3pt]
v[g\ls{\hh\hp\hm}-g\ls{\hh\ha\ha}]&=&
    -2(\mhpm^2-\mha^2)\cba\,.
\eeqa

\section{Four-Higgs vertices in the two-Higgs doublet model}
      \label{app:E}

In this Appendix, we list the Feynman rules for the four-point Higgs
interaction in the most general CP-conserving
two-Higgs doublet extension of the Standard Model.
Recalling that $\mathcal{L}_{\rm int} \ni -\mathcal{V}$,
the Feynman rules are obtained from the scalar 
potential\footnote{Note, {\it e.g.}, that the term 
proportional to $\hl\ha\hp\gm$ in $\mathcal{V}$ 
corresponds to $\hp$ and $\gm$ directed {\it into} the vertex, {\it etc.}} 
by multiplying the corresponding coefficients of $\mathcal{V}$ by $-i$
times the appropriate symmetry factor.
We find it convenient to write the terms of the potential that are
quartic in the Higgs fields as a sum of two pieces: 
$\mathcal{V} \ni \mathcal{V}_A+\mathcal{V}_B$, where
$\mathcal{V}_A$ depends explicitly on $\beta-\alpha$ 
and $\mathcal{V}_B$ is independent of $\beta-\alpha$.
To obtain the four-Higgs couplings in terms of $\beta-\alpha$ and
the invariant coupling parameters, we
insert \eqs{abbasis}{Hbasis} into \eq{pothbasis}, and
identify the terms that are quartic in the Higgs boson fields.
For completeness, the quartic interaction terms
involving the Goldstone bosons ($\gpm$ and $\go$) are also listed.
The end result is:
\beqa\label{v4a}
&&8\mathcal{V}_A=
\nonumber \\[3pt] &&
\hl^{4} \Bigl[\lam\sbma^2(3\cbma^2+1)
-4\lamhat\cbma\sbma^3-2(2\lamA-\lamT)\cbma^2\sbma^2 
-4\lamU\cbma^3\sbma+\lamV\cbma^4\Bigr]
\nonumber \\[3pt] && 
+ 4\hl^{3} \hh \Bigl[\lam\sbma\cbma(3\cbma^2-1)-\lamhat\sbma^2(4\cbma^2-1)
-(2\lamA-\lamT)\sbma\cbma(\cbma^2-\sbma^2)
\nonumber \\ &&\qquad\qquad\qquad\qquad\qquad\qquad
+\lamU\cbma^2(4\sbma^2-1)-\lamV\sbma\cbma^3\Bigr]
\nonumber \\[3pt] 
&& +2 \hl^{2}
\hh^{2} \Bigl[\lam(2-9\sbma^2\cbma^2)-6(\lamhat-\lamU)\sbma\cbma(\cbma^2
-\sbma^2)
\nonumber \\ &&\qquad\qquad\qquad\qquad\qquad\qquad
-(2\lamA-\lamT)(1-6\sbma^2\cbma^2)+3\lamV\sbma^2\cbma^2\Bigr]
\nonumber \\[3pt] 
&&+4 \hl \hh^{3}
\Bigl[\lam\sbma\cbma(3\sbma^2-1)+\lamhat\cbma^2(4\sbma^2-1)
+(2\lamA-\lamT)\sbma\cbma(\cbma^2-\sbma^2)
\nonumber \\ &&\qquad\qquad\qquad\qquad\qquad\qquad
-\lamU\sbma^2(4\cbma^2-1)-\lamV\cbma\sbma^3\Bigr]
\nonumber \\[3pt] 
&&+\hh^{4} \Bigl[\lam\cbma^2(3\sbma^2+1)+4\lamhat\cbma^3\sbma
-2(2\lamA-\lamT)\cbma^2\sbma^2+4\lamU\cbma\sbma^3+\lamV\sbma^4\Bigr]
\nonumber \\[3pt] 
&&+2 \hl^{2} \ha^{2} \Bigl[\lamT\sbma^2-2\lamU\sbma\cbma+\lamV\cbma^2\Bigr]
\nonumber \\[3pt] 
&&+4\hl^{2} \ha \go \Bigl[2(\lam-\lamA)\sbma\cbma-\lamhat\sbma^2-
\lamU\cbma^2\Bigr]
\nonumber \\[3pt] 
&&+2 \hl^{2} \go^{2} \Bigl[\lam\sbma^2-2\lamhat\sbma\cbma+\lamT\cbma^2\Bigr]
\nonumber \\[3pt] 
&&+4 \hl^{2}\hp \hm \Bigl[(\lamT+\lamF)\sbma^2-2\lamU\sbma\cbma
+\lamV\cbma^2\Bigr]
\nonumber \\[3pt] 
&&+4 (\hl^{2}\hp \gm+\hl^{2}\hm\gp) \Bigl[(2\lam-2\lamA-\lamF)\sbma\cbma
-\lamhat\sbma^2-\lamU\cbma^2\Bigr]
\nonumber \\[3pt] 
&&+4 \hl^{2} \gp \gm \Bigl[\lam\sbma^2-2\lamhat\sbma\cbma
+(\lamT+\lamF)\cbma^2\Bigr]
\nonumber \\[5pt] 
&&+4 \hl \hh \ha^{2}
\Bigl[(\lamT-\lamV)\sbma\cbma-\lamU(\cbma^2-\sbma^2)\Bigr]
\nonumber \\[3pt] 
&&+8 \hl \hh \ha \go \Bigl[(\lam-\lamA)(\cbma^2-\sbma^2)-
(\lamhat-\lamU)\sbma\cbma)\Bigr]
\nonumber \\[3pt] 
&&+4 \hl \hh \go^{2} \Bigl[(\lam-\lamT)\sbma\cbma
-\lamhat(\cbma^2-\sbma^2)\Bigr]
\nonumber \\[3pt] 
&&+8 \hl \hh \hp \hm \Bigl[(\lamT-\lamV+\lamF)\sbma\cbma
-\lamU(\cbma^2-\sbma^2)\Bigr]
\nonumber \\[3pt] 
&&+4( \hl \hh \hp\gm+\hl\hh\hm\gp)
\Bigl[(2\lam-2\lamA-\lamF)(\cbma^2-\sbma^2)
-2(\lamhat-\lamU)\sbma\cbma\Bigr]
\nonumber \\[3pt] 
&&+8 \hl \hh \gp \gm
\Bigl[(\lam-\lamT-\lamF)\sbma\cbma-\lamhat(\cbma^2-\sbma^2)\Bigr]
\nonumber \\[3pt] 
&&+2 \hh^{2} \ha^{2} \Bigl[\lamT\cbma^2+2\lamU\sbma\cbma+\lamV\sbma^2\Bigr]
\nonumber \\[3pt] 
&&+4 \hh^{2} \ha \go \Bigl[2(\lamA-\lam)\sbma\cbma-\lamhat\cbma^2
-\lamU\sbma^2\Bigr]
\nonumber \\[3pt] 
&&+2 \hh^{2} \go^{2} \Bigl[\lam\cbma^2+2\lamhat\sbma\cbma
+\lamT\sbma^2\Bigr]
\nonumber \\[3pt] 
&&+4 \hh^{2} \hp \hm
\Bigl[(\lamT+\lamF)\cbma^2+2\lamU\sbma\cbma+\lamV\sbma^2\Bigr]
\nonumber \\[3pt] 
&&+4(\hh^{2}\hp \gm+ \hh^{2}\hm\gp) \Bigl[(2\lamA-2\lam+\lamF)\sbma\cbma
-\lamhat\cbma^2-\lamU\sbma^2\Bigr]
\nonumber \\[3pt] 
&&+4 \hh^{2} \gp \gm \Bigl[\lam\cbma^2+2\lamhat\sbma\cbma
+(\lamT+\lamF)\sbma^2\Bigr]
\nonumber \\[3pt]
 &&+4 i \Bigl[\hl \ha \hp \gm - \hl\ha\hm\gp
+\hh\go\hp\gm-\hh\go\hm\gp\Bigr] \lamF\sbma 
\nonumber \\ && -4 i \Bigl[\hl\go\hp\gm-\hl\go\hm\gp
-\hh\ha\hp\gm+\hh\ha\hm\gp\Bigr]\lamF\cbma\,,
\eeqa
and 
\beqa
8\mathcal{V}_B &=&\lamV (\ha^{4}+4\ha^2\hp\hm+4\hp\hm\hp\hm)
\nonumber\\
&&\!\!\!\!\!\!\!\!  - 4\lamU (\ha^{3} \go+\ha^2\hp\gm
+\ha^2\hm\gp+2\ha\go\hp\hm+2\hp\hm\hp\gm+2\hp\hm\hm\gp)\nonumber\\
&&\!\!\!\!\!\!\!\!
+2\left[2(\lam -\lamA)+\lamT\right] \ha^{2} \go^{2} 
\nonumber\\
&& \!\!\!\!\!\!\!\!
  +4 (\lamT+\lamF)(\ha^{2} \gp \gm +\go^2\hp\hm)
\nonumber\\
&&\!\!\!\!\!\!\!\! 
  -4 \lamhat (\ha \go^{3} +2\ha\go\gp\gm+\go^2\hp\gm+\go^2\hm\gp
+2\hp\gp\gm\gm+2\hm\gm\gp\gp)\nonumber \\
&&\!\!\!\!\!\!\!\!
  +4\left[2(\lam-\lamA)-\lamF \right](\ha \go \hp\gm +\ha\go\hm\gp)
\nonumber\\
&&\!\!\!\!\!\!\!\! 
+\lam (\go^{4}+4\go^2\gp\gm+4\gp\gm\gp\gm) 
\nonumber\\
&&\!\!\!\!\!\!\!\!
+4(\lam-\lamA)(\hp\hp\gm\gm+\hm\hm\gp\gp)
\nonumber\\
&&\!\!\!\!\!\!\!\! 
+ 8(\lam-\lamA+\lamT)\hp\hm\gp\gm\,.
\label{v4b}
\eeqa

The quartic Higgs couplings are now easily obtained by including the
appropriate symmetry factors.  For example, the
$\hl^4$ and $\hh^4$ couplings are given by
\beqa
g_{\hl\hl\hl\hl}&=& -3\bigl[\lam\sbma^2(1+3\cbma^2)-4\lamhat\cbma\sbma^3
            -2(2\lamA-\lamT)\cbma^2\sbma^2 \nonumber \\
&&\qquad\qquad -4\lamU\cbma^3\sbma+\lamV\cbma^4\bigr]
\,, \label{invfourhiggs1} \\
g_{\hh\hh\hh\hh}&=& -3\bigl[\lam\cbma^2(1+3\sbma^2)+4\lamhat\cbma^3\sbma
            -2(2\lamA-\lamT)\cbma^2\sbma^2 \nonumber\\
&&\qquad\qquad +4\lamU\cbma\sbma^3+\lamV\sbma^4\bigr]\,.\label{invfourhiggs2}
\eeqa
Note the first appearance of physical observables that depend on $\lamV$. 

Let us denote the Feynman rule for the $ABCD$ vertex by
$ig\ls{ABCD}$.  In the approach to the
decoupling limit, the four-Higgs vertices simplify
considerably as exhibited in Table~\ref{hhhhdecouptab}.  Here, we have
listed all couplings in the form:
\beq \label{gabcd}
g_{ABCD}= (X_{ABCD}+Y_{ABCD}\,\cbma)\,,
\eeq
where the coefficients $X$ and $Y$ are given in terms of various
linear combinations of the invariant coupling parameters.
Note that the terms contained in $\mathcal{V}_B$ are not affected
by the decoupling limit since these terms are independent of $\beta-\alpha$.

\begin{table}[t!]
\begin{center}
\caption{Four-Higgs vertex Feynman rules in the approach to the decoupling
limit are given by $ig_{ABCD}=i(X_{ABCD}+Y_{ABCD}\,\cbma)$, where the
coefficients $X$ and $Y$ are listed below. 
The rules for $\ha\ha\ha\ha$, $\ha\ha\hp\hm$,
and $\hp\hm\hp\hm$ are exact (since they are independent of 
$\beta-\alpha$).
\label{hhhhdecouptab}}

\vspace{.2in}
\begin{tabular}{|l|c|c|} \hline
$ABCD$ & $X_{ABCD}$ & $Y_{ABCD}$ \rule{0in}{3ex} \\[1ex] \hline
$\hl\hl\hl\hl$ & $-3\lam$  &  
                 \rule{0in}{3ex}  $12\lamhat$  \rule{0in}{3ex} \\[4pt]
$\hl\hl\hl\hh$ & $-3\lamhat$ &   $3(\lam+\lamT-2\lam_A)$ \\[4pt]
$\hl\hl\hh\hh$ & $2(\lam_A-\lam)-\lamT$ &   $6(\lamU-\lamhat)$ \\[4pt]
$\hl\hl\ha\ha$ & $-\lamT$ &   $2\lamU$ \\[4pt]
$\hl\hl\hp\hm$ & $-\lamT-\lamF$ &   $2\lamU$ \\[4pt]
$\hl\hh\hh\hh$ & $-3\lamU$ &  $3(\lamV-\lamT)+6(\lamA-\lam)$ \\[4pt]
$\hl\hh\ha\ha$ & $-\lamU$ &  $-\lamT+\lamV$ \\[4pt]
$\hl\hh\hp\hm$ & $-\lamU$ &  $\lamV-\lamT-\lamF$ \\[4pt]
$\hh\hh\hh\hh$ & $-3\lamV$ &  $-12\lamU$ \\[4pt]
$\hh\hh\ha\ha$ & $-\lamV$ &  $-2\lamU$ \\[4pt]
$\hh\hh\hp\hm$ & $-\lamV$ &  $-2\lamU$ \\[4pt]
$\ha\ha\ha\ha$ & $-3\lamV$ &  $0$ \\[4pt]
$\ha\ha\hp\hm$ & $-\lamV$ &  $0$ \\[4pt]
$\hp\hm\hp\hm$ & $-2\lamV$ &  $0$ \\[1ex] \hline
\end{tabular}
\end{center}
\end{table}

The four-Higgs couplings can be rewritten in terms of 
$\lam_1,\cdots,\lam_7$, $\alpha$ and $\beta$.  The resulting
expressions are generally more complex, with a few notable exceptions.
For example, the quartic couplings 
in $\mathcal{V}_A$ that depend only on $\hl$ and $\hh$
are independent of $\beta$
\beqa \label{v4ap}
\mathcal{V}_A&\ni& \phm\eighth\,
\hl^{4} \Bigl[\lam_1 \sa^{4}+\lam_2 \ca^{4}+\half\lamtil
  \stwoa^2-2\stwoa( \lam_6  \sa^{2}+\lam_7 \ca^{2}) \Bigr]
\nonumber 
\\[3pt] && 
+ \half\,\hl^{3} \hh \Bigl[\half\stwoa(-\lam_1 \sa^{2}+\lam_2 \ca^{2} -\lamtil
  \ctwoa)+\lam_6\sa\sthreea+\lam_7 \ca\cthreea \Bigr]
\nonumber 
\\[3pt] && 
+\quarter\, \hl^{2}
\hh^{2} \Bigl[\nicefrac{3}{4}\stwoa^2(\lam_1+\lam_2 -2\lamtil)+\lamtil 
-3\stwoa\ctwoa(\lam_6-\lam_7)\Bigr]
\nonumber \\[3pt] 
&&+\half\, \hl \hh^{3}
\Bigl[\half\stwoa(-\lam_1 \ca^2+\lam_2\sa^{2}+\lamtil\ctwoa)
+\lam_6\ca\cthreea+\lam_7\sa\sthreea\Bigr]
\nonumber \\[3pt] 
&&+\eighth\,\hh^{4} \Bigl[\lam_1 \ca^{4}+\lam_2 \sa^{4}+\half \lamtil
  \stwoa^2+2\stwoa(\lam_6 \ca^{2}+ \lam_7 \sa^{2})\Bigr]\,,
\eeqa
and in this form these results are somewhat simpler than
the corresponding expressions 
in terms of the invariant coupling parameters given in \eq{v4a}.
One can check that the latter can be obtained from \eq{v4ap} by
rotating to the Higgs basis (see discussion in Appendix~\ref{app:C2}).
That is, in \eq{v4ap}, let $\alpha\to\alpha-\beta$, $\lam_1\to\lam$,
$\lam_2\to\lam_V$, $\lamtil\to 2(\lam-\lamA)+\lamT$, $\lam_6\to
-\lamhat$ and $\lam_7\to -\lamU$ [{\it cf.}~\eq{pothbasis}].

One can also express the four-Higgs vertices in terms of the Higgs
masses by using \eqs{inverseA}{inverseD}.  For example~\cite{boudjema},
\beqa
\!\!\!\!\!\!\!\!
g_{\hl\hl\hl\hl}&=& -3\left[{\mhl^2\over
    v^2}\left(\sbma-{\cbpa\cbma^2\over\sb\cb}\right)^2
+{\mhh^2\over v^2}\left({\sa\ca\cbma\over\sb\cb}\right)^2 \right.
\nonumber \\[5pt]
&&\left.\qquad
-{\mha^2+\lam_5 v^2\over v^2}\left({\cbpa\cbma\over\sb\cb}\right)^2
-{2(\lam_6\sa^2+\lam_7\ca^2)\cbma^2\over\sb\cb}\right]\,.
\eeqa
Note that the
decoupling limit result [\eq{hhhhdecoup}] follows trivially,
after using \eqns{decouplimits1}{decouplimits3} 
to obtain the $\mathcal{O}(\cbma)$
correction.  Expressions for other four-Higgs couplings in
terms of the Higgs masses can be found in \Ref{boudjema} (see also
\Ref{Dubinin:1998nt} for the case of $\lam_6=\lam_7=0$
and \Ref{santosetal} for other special cases).  However, in the most
general case, such expressions are less useful.


\end{document}